\documentclass[a4paper,fleqn,preprint]{cas-dc}

\usepackage[numbers,sort&compress]{natbib}

\def\tsc#1{\csdef{#1}{\textsc{\lowercase{#1}}\xspace}}
\tsc{WGM}
\tsc{QE}

\usepackage[english]{babel}
\usepackage[utf8]{inputenc}
\usepackage[T1]{fontenc}
\usepackage{tgtermes}
\usepackage{cleveref}
\crefname{subsection}{subsection}{subsections}
\usepackage{todonotes}
\usepackage{placeins}

\usepackage{multirow}
\usepackage{siunitx}
\usepackage{graphicx}
\usepackage{dcolumn}

\DeclareSIUnit\angstrom{\text {Å}}

\usepackage[caption=false]{subfig}
\usepackage{bm}

\DeclareSIUnit{\atmospheric}{atm}

\begin{document}
\let\WriteBookmarks\relax
\def\floatpagepagefraction{1}
\def\textpagefraction{.001}

\shorttitle{Characterising anisotropic diffusion of complex fluids in confined environments}    

\shortauthors{H\"ollring et al.}  

\title [mode = title]{Anisotropic molecular diffusion in confinement II: \\A model for structurally complex particles applied to transport in thin ionic liquid films}

\author[1]{Kevin H\"ollring}[orcid=0000-0002-9497-3254]
\credit{Conceptualization, Methodology, Software, Formal analysis, Investigation, Writing - Original Draft, Visualization, Data Curation}

\author[1]{Andreas Baer}[orcid=0000-0001-7943-8846]
\credit{Simulation, Contribution to Methods sections}

\author[2,1]{Nataša Vučemilović-Alagić}[orcid=0000-0001-5841-7181]
\credit{Simulations, Writing - Original Draft (methods)}

\author[2]{David M. Smith}[orcid=0000-0002-5578-2551]
\credit{Design of MD simulations}

\author[1,2]{Ana-Sunčana Smith}[orcid=0000-0002-0835-0086]
\credit{Conceptualization, Investigation, Writing - Review \& Editing, Project coordination, Funding acquisition, Resources, Data Curation}
\ead{ana-suncana.smith@fau.de, asmith@irb.hr}

\cormark[1]
\fnmark[1]

\affiliation[1]{organization={PULS Group, Institute for Theoretical Physics, FAU Erlangen-N\"urnberg},
addressline={Cauerstra\ss{}e 3},
postcode={91058}, 
city={Erlangen},
country={Germany}}

\affiliation[2]{organization={Group of Computational Life Sciences, Department of Physical Chemistry, Ru\dj{}er Bo\v{s}kovi\'{c} Institute},
addressline={Bijeni\v{c}ka 54},
city={Zagreb},
postcode={10000},
country={Croatia}}

\cortext[1]{Corresponding author}
\fntext[1]{Tel: +49 91318570565; Fax: +49 91318520860}

\date{\today}

\begin{abstract}
\indent
\paragraph{Hypothesis:}
Diffusion in confinement is an important fundamental problem with significant implications for applications of supported liquid phases. 
However, resolving the spatially dependent diffusion coefficient, parallel and perpendicular to interfaces, has been a standing issue and for objects of nanometric size, which structurally fluctuate on a similar time scale as they diffuse, no methodology has been established so far.   
We hypothesise that the complex, coupled dynamics can be captured and analysed by using a model built on the $2$-dimensional Smoluchowski equation and systematic coarse-graining.
\paragraph{Methods and simulations:}
For large, flexible species, a universal approach is offered that does not make any assumptions about the separation of time scales between translation and other degrees of freedom. 
The method is validated on Molecular Dynamics simulations of bulk systems of a family of ionic liquids with increasing cation sizes where internal degrees of freedom have little to major effects. 
\paragraph{Findings:}
 After validation on bulk liquids, where we provide an interpretation of two diffusion constants for each species found experimentally, we clearly demonstrate the anisotropic nature of diffusion coefficients at interfaces.
 Spatial variations in the diffusivities relate to interface-induced structuring of the ionic liquids.  
 Notably, the length scales in strongly confined ionic liquids vary consistently but differently at the solid-liquid and liquid-vapour interfaces.
\end{abstract}

\begin{graphicalabstract}
\includegraphics{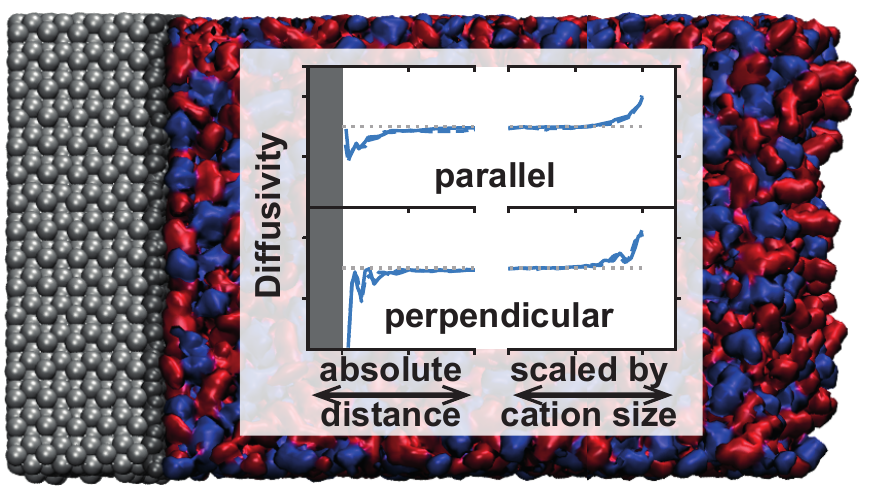}
\end{graphicalabstract}


\maketitle
\section{Introduction}
Molecular transport in strong confinement is a multi-scale problem \cite{lhermerout2018ionic,gebbie2017long}. 
It is particularly acute when there is no clear spatial length separation between the size of molecules within the fluid, the confinement length-scale, and the length scale of the interaction potential between the diffusing particle and the solid phase \cite{smith2017switching,raviv2004fluidity}. Namely, close to solid-liquid (SL)\footnote{\textbf{Abbreviations:} Ionic Liquid (IL), 
Mean Square Displacement (MSD), 
Simple Particle Model (SPM), 
Simple Particle Model accounting for diffusion (SPM+d), 
Extensive Particle Model (EPM), 
Local Width Reduction (LWR), 
Solid-Liquid (SL), 
Liquid-Vacuum (LV), 
Solid-Liquid-Vacuum (SLV), 
Interface Normal Number Density (INND), Partial Differential Equation (PDE)} \cite{tournassat2016molecular,somers2013review,perez2017scaling} and liquid-vapour (LV) \cite{wertheim1976correlations,ball2003keep,lovett1976structure} interfaces, solvent layering affects the potential of mean force between the diffusing particle and the interface itself.
This can yield a higher level of organisation at the interface compared to the bulk liquid, in turn affecting diffusion into and within the interface layers \cite{lovett1976structure,kuo2004ab,vucemilovic2019insights,wertheim1976correlations,steinruck2011surface,tsuzuki2012factors}.

The consequence of the interface-induced change in the free energy landscape are modifications of the preferred conformations of the molecules within the fluid \cite{lhermerout2018ionic,perkin2012ionic,ravichandran1999anisotropic}. 
In the context of transport, different average molecular shapes at the interface and in the bulk should be associated with different effective hydrodynamic radii \cite{han2006brownian,pande2015forces,ravichandran1999anisotropic}. 
Furthermore,  structural fluctuations of the diffusing molecules \cite{bresme2007nanoparticles}, induced by frequent interactions with neighbours and external potentials, may be coupled to translations on certain time scales already in bulk liquids, as they occur under the influence of the same interactions that affect simple diffusion \cite{zwanzig1970hydrodynamic}.
Close to interfaces, both amplitudes and timescales at which fluctuations occur can be strongly modified \cite{ball2003keep,lovett1976structure,wilson1987molecular}, which may be particularly important in molecular liquids where the constituents itself are sizable molecules. 
All these effects can both hinder or promote diffusive transport  close to interfaces compared to bulk configurations and introduce anisotropic mobility parallel and perpendicular to the confining surface \cite{mittal2008layering,fernandez2004self,nordanger2022anisotropic,mittal2008layering}. 
The extent of this coupling will depend on the specific properties of the materials involved, preferred particle conformations and the direction of motion \cite{han2006brownian,pande2015forces}.

The discussed phenomena are particularly important for ionic liquids (ILs) \cite{marion2021water,fedorov2014ionic,somers2013review,tsuzuki2012factors,minami2009ionic,canongia2006nanostructural,smith2016electrostatic}.
ILs are molten salts with very low vapour pressure that remain in the liquid state even at room temperatures due to strong Coulomb interactions and asymmetric ion sizes \cite{fajardo2017water,tournassat2015ionic,fedorov2014ionic,kornyshev2007double,canongia2006nanostructural,rebelo2005critical,perkin2012ionic}. 
Their structural properties, and in particular those belonging to the 1-alkyl-3-methylimidazolium bis(trifluoromethylsulfonyl)imide [C$_1$C$_n$Mim] series of cations combined with the bistriflimide [Ntf$_2$] anions, have been extensively studied \cite{merlet2014electric,salanne2016efficient,merlet2013computer,futamura2017partial,padua2007molecular,canongia2006nonpolar,tariq2012surface,canongia2006nanostructural,smith2016electrostatic,smith2017switching}. 
The cations [C$_1$C$_n$Mim] can vary in chain length from $n=2$ being comparable in size to the anion (ca. \SI{0.4}{\nano\meter}) to $n=10$, which is about $\SI{1.3}{\nano\meter}$ long, which leads to strong coupling close to interfaces. 
On hydroxylated alumina their adsorption is mainly governed by the formation of hydrogen bond networks \cite{segura2013adsorbed, vucemilovic2019insights}. 
The formed films are highly stable, with up to 8 or even 9 solvation layers forming on top of a solid \cite{lhermerout2018ionic,perez2017scaling,vucemilovic2019insights,brkljaca2015complementary, horn1988double,atkin2007structure}, with cations arranging the polar heads parallel to the surface \cite{steinruck2011surface,fedorov2008ionic}. 
On the liquid vacuum interface, the alkyl chains organise into a hydrophobic layer, positioning the axis of the cation perpendicular to the interface \cite{steinruck2011surface,sobota2010toward}.

\begin{figure}
\includegraphics[width=\columnwidth]{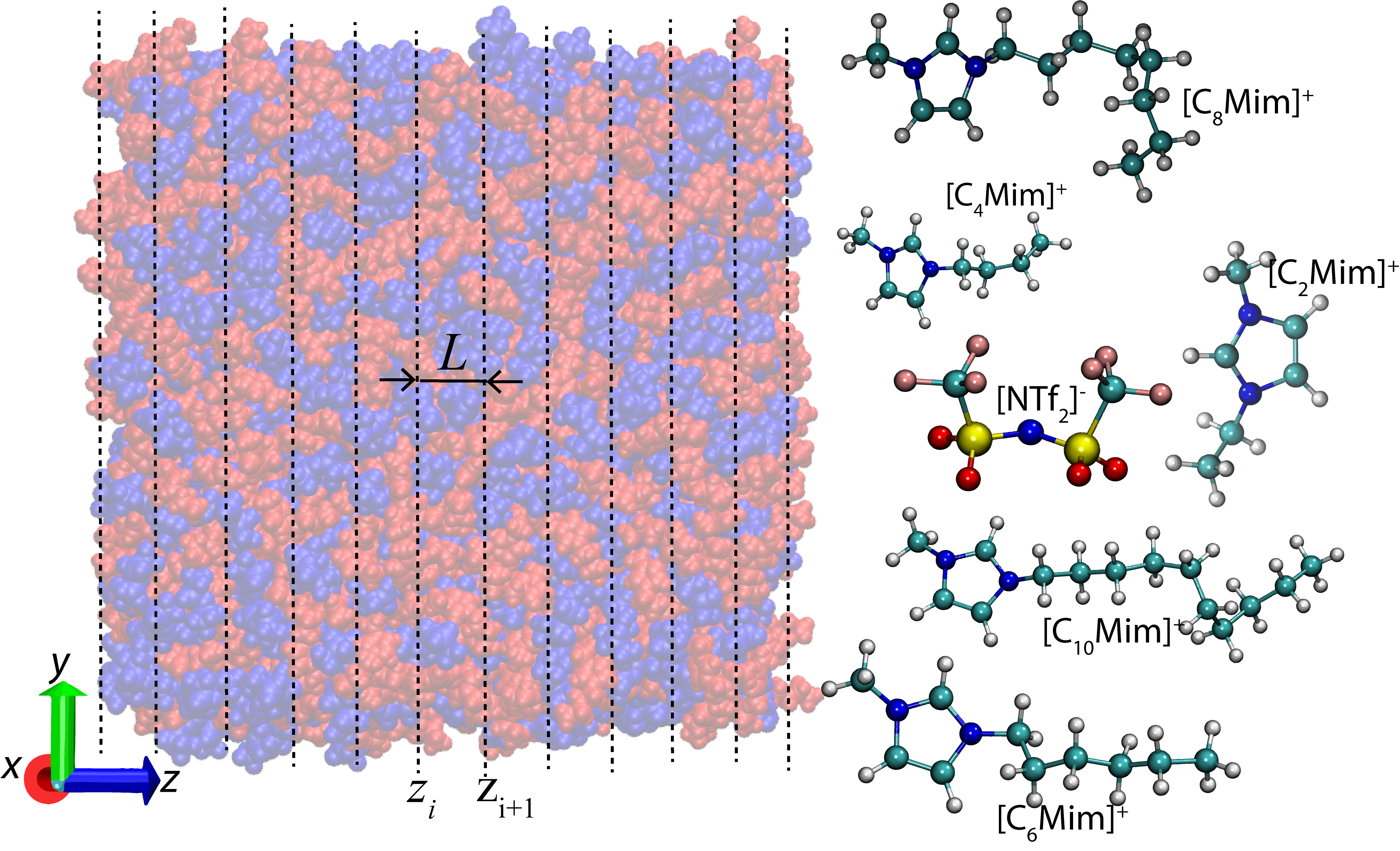}
\caption{\textbf{Bulk IL system slicing and molecular structure:}\label{fig:simulation_samples} 
The simulation box for a bulk IL ($1000$ ion pairs) sliced into slabs for the analysis of diffusion based on particle lifetime, together with a series of imidazolium-based cations [C$_n$Mim]$^+$ of different alkyl chain lengths ($n = 2,4,6,8,10$) and the [NTf$_2$]$^-$ anion. \textit{(Figure adapted from \cite{vuvcemilovic2021computational})}}
\end{figure}

Characteristics of ILs' diffusive transport, especially in confinement are still highly debated \cite{araque2015diffusion}. 
On one hand, the difficulties to understand this system are rooted in the size of the ions and their complex correlations.
But on the other hand, the methods available to resolve diffusing in confinement are limited. 
The commonly used approach to address these challenges are molecular dynamics (MD) simulations \cite{fedorov2008towards,merlet2014electric,salanne2011polarization,kondrat2014accelerating,salanne2012including,salanne2015simulations,padua2007molecular,canongia2006nanostructural,canongia2004modeling,canongia2004molecular, horstmann2022structural}.
However, in this case, the transport coefficients need to be extracted from recorded trajectories.

In a prequel to this work, we extensively discussed the available methodologies to calculate spatially dependent diffusion coefficients from MD simulations \cite{spm_pub}.
We devised the Simple Particle Model (SPM) for the analysis of interface-normal diffusivity based on local lifetime statistics of simple point-like particles. 
We demonstrated in water filled pores that the SPM accurately captures the diffusivity profile despite the presence of a statistical drift close to the interfaces.
This was confirmed via the extension of the basic model that accounts for the influence of drift induced by a gradient in the effective interaction potential (referred to as SPM+d). 
As a result, we found that the diffusivity of water in a pore is highly anisotropic with the perpendicular component showing a somewhat unexpected, non-monotonous behaviour.

Applying this new methodology to more complex liquids like ILs is a natural step forward. 
However, ILs clearly violate the basic premises of the SPM as well as those of all other currently available approaches, which we demonstrate herein. 
Therefore, to resolve anisotropic, and spatially dependent diffusion coefficients of extended, flexible molecules we expand on the previous theoretical analysis.
In doing so, we develop the so-called Extensive Particle Model (EPM) to analyse interface-normal diffusivity in complex fluids. 
Our extended model explicitly captures the internal fluctuations of extensive particles and their influence on the observed lifetime statistics, which is a unique advantage of our approach compared to existing methodologies. 
Consequently, this allows us to fully resolve the diffusion profiles of cations and anions within IL films and study the effect of the ion size and density correlations on diffusive transport in confinement.

\section{Bulk ILs and the dawn of the Simple Particle Model}

We first check the applicability of the established SPM approach for point-like particles \cite{spm_pub} on bulk IL systems. 
To apply the SPM, the system is cut into parallel slices of thickness $L$ (\cref{fig:simulation_samples}).
The SPM then establishes a link between the mean diffusion coefficient orthogonal to the cuts $D_\perp$ and the mean expected particle lifetime within a slice based on the drift-free Smoluchowski equation \cite{belousov2022statistical}. 
Applying appropriate boundary conditions yields 
\begin{align}
\left\langle D_\perp  (z)\right\rangle  =c\times \frac{L^2}{\tau}.\label{eq:D_relation_basic_shape}
\end{align}
with a coefficient $c_\mathrm{B}=1/12$ for bulk-like slabs (i.e. particles can leave to either side) and $c_{LV}=1/3$ for LV-like slabs (i.e. particles can only escape to one side) under the assumption of a constant background potential. 

An extension to the SPM that accounts for the effect of a constant drift $\mu$ across the slab, is referred to as SPM+d \cite{spm_pub}.
For bulk-like slices, it results in
\begin{align}
\left\langle D_{\perp+d}  (z)\right\rangle  = \frac{1}{12}\times K\left(\frac{\mu L}{D}\right) \frac{L^2}{\tau},\label{eq:D_spmd}
\end{align}
where $K(\gamma)$ denotes a correction factor based on the relative amplitude $\gamma=\mu L/D$.
For small values of $\gamma\leq 2$, the influence of this correction is minimal \cite{spm_pub}.

To explore the SPM in imidazolium-based  ILs, we first perform MD simulations of bulk systems (see appendix \ref{app:simulation_methods} for methodological details). 
Each bulk IL consists of a thousand small $[NTf_2]^-$ anions and the same number of [C$_n$Mim]$^+$ cations, where the alkyl chain length is systematically increased ($n = 2,4,6,8,10$), placed in a box with periodic boundary conditions. 
Diffusion data originate from a $\SI{100}{\nano\second}$ long production run. 
The expectation is that with increasing cation size the performance of the SPM worsens.

\begin{table}
\caption{\label{tab:correction_model_benchmark_spm}%
\textbf{Testing the performance of the SPM in bulk ILs:}
ILs with $n$ carbon atoms in the alkyl chain of the cation are investigated.
Reference bulk self-diffusion coefficients (MSD) for cations and anions (superscript $+/-$) are obtained from MD trajectories using the full-system analysis of mean square displacements with \texttt{gmx msd}. 
Results for the SPM are obtained through \cref{eq:D_relation_basic_shape} with $L=\SI{1}{\nano\meter}$.
Furthermore, $\mathrm{SPM}_\mathrm{Err}$ denotes the error of $D= D_\mathrm{SPM}$ relative to the MSD value as defined in \cref{eq:msd_err}.  
Experimental results (EXP) are adapted from Tokuda et \emph{al.} \cite{tokuda2005physicochemical}  as described in appendix~\ref{app:rtils_experimental}.  
All diffusion constants are in units of $\SI{e-7}{\centi\meter\squared\per\second}$.}
\begin{tabular}{c|cccc}
\toprule
&\multicolumn{4}{c}{$D^+$}\\
$n$	& MSD & SPM & SPM$_{\mathrm{Err}}$ & EXP \cite{tokuda2005physicochemical}\\
\midrule
2  	& 
$3.5\pm 0.1$ & $6.0\pm 0.2$ &$\SI{71}{\percent}$ & $5.3 \pm 2.6$\\
4	& 
$2.6\pm 0.1$ & $4.8\pm 0.1$ & $\SI{85}{\percent}$ &  $3.0\pm 1.4$\\
6	& 
$1.4\pm 0.1$ & $2.8\pm 0.1$ & $\SI{100}{\percent}$ & $1.8 \pm 1.2$ \\
8	& 
$0.8\pm 0.1$ & $2.0\pm 0.1$ & $\SI{150}{\percent}$ & $1.3 \pm 1.4$\\
10	& 
$0.7\pm 0.1$ & $1.7\pm 0.1$ & $\SI{142}{\percent}$ & --\\
\midrule
&\multicolumn{4}{c}{$D^-$}\\
$n$	& MSD & SPM & SPM$_{\mathrm{Err}}$ & EXP \cite{tokuda2005physicochemical}\\
\midrule
2  	& 
$2.0\pm 0.1$ & $3.8\pm 0.1$ & $\SI{90}{\percent}$ & $3.3 \pm 1.4$\\
4	& 
$1.7 \pm 0.1$ & $3.5\pm 0.1$ & $\SI{106}{\percent}$& $2.4 \pm 2.2$\\
6	& 
$1.1 \pm 0.1$ & $2.4\pm 0.1$ & $\SI{120}{\percent}$ & $1.7 \pm 1.0$ \\
8	& 
$0.7 \pm 0.1$ & $1.9\pm 0.1$ & $\SI{170}{\percent}$ & $1.3 \pm 0.9$\\
10	& 
$0.6 \pm 0.1$ & $1.7\pm 0.1$ & $\SI{180}{\percent}$ &  --\\
\bottomrule
\end{tabular}
\end{table}

However, in order to validate the SPM, reference bulk values of diffusion coefficients must be established. 
Therefore, we first perform a standard analysis of diffusivities based on the Mean Square Displacement (MSD) of the ions. 
The results in \cref{tab:correction_model_benchmark_spm} show that our model system agrees well with experimentally measured anion and cation diffusion constants at relevant temperatures, both in terms of absolute diffusivities, as well as the apparent cationic transference number \cite{tokuda2005physicochemical}.
Like in experiments (see \cref{tab:correction_model_benchmark_spm}), we find that cations have higher self-diffusion coefficients than anions, even at large cation sizes. 
However, with increasing $n$, the difference between $D^+$ and $D^-$ becomes smaller, such that the apparent transference number becomes close to $0.5$ for $n=8$. 
We conclude that our force field produces transport behaviour comparable to experimental observations as supported by the results of the established method (Einstein/MSD) which is certainly applicable to the bulk system.

Applying the SPM to the pure liquid bulk ILs using $L=\SI{1}{\nano\meter}$ results in drastic deviations from the reference diffusivities (see \cref{tab:correction_model_benchmark_spm}). 
Here MSD-based and not experimental data are used for the comparison as the benchmark needs to be executed on data shared with the reference. 
This specifically validates the performance of the analysis method and not that of the chosen force field.
In the comparison of the SPM and MSD approaches we find large deviations already for systems with the smallest considered cation, due to strong anion-cation interactions. 
The results further deviate as $n$ increases from $2$ to $10$. 
Specifically, for cations the SPM error 
\begin{align}
    \frac{|D_\mathrm{MSD}-D|}{D_\mathrm{MSD}} \label{eq:msd_err}
\end{align} 
of the SPM increases from $\SI{75}{\percent}$ to $\SI{160}{\percent}$. 
The large and further increasing errors in the diffusivities of cations may be attributed to the gradual prevalence of the fluctuations of their centre of mass due to internal degrees of freedom being affected by inter-molecular interactions. 
These interactions coincide with increasing local structuring in imidazolium-based ILs as a result of increasing cation chain length, as has been predicted by simulations and theory but also been observed experimentally \cite{iwata2007local,kirchner2015ion}.
The fluctuations of the centre of mass shorten the mean lifetime of cations within the slab, such that the SPM more and more overestimates the diffusivity even if the viscosity of the IL decreases. 

The errors are even larger for anion diffusion.  
In this case, the fluctuations of internal degrees of freedom play little role.
Instead, due to strong attractive interactions, positional fluctuations of the anions are strongly correlated with those of the cations, even for $n=2$. 
These correlations increase with the cation size due to stronger ionic coordination appearing in the IL, which have also been observed in experimental studies \cite{koddermann2006ion,katoh2008ion,katsuta2008ion,wang2020microstructural}. 
Namely, [C$_n$mim][Ntf$_2$] ILs were found to develop continuous, relatively dense \emph{ionic} regions and --- for shorter chains  --- islands of non-polar regions, the latter growing with $n$. 
Above $n=6$ the non-polar domains percolate into a second continuous region overall forming a bi-continuous fluid phase \cite{rocha2011high}. 
At this point, translational diffusion of the cation is basically the same as that of the anion due to strong correlations between the delocalised charges that the ions possess. 
This coupling induces faster excursions of the anion from the slab than expected, contributing to the error of the SPM. 

These results demonstrate that the SPM is not suitable for the analysis of diffusion in ILs.
Importantly, our results point to a general shortcoming of Markov-State-Model-style approaches for determining diffusion coefficients.  
These models cannot account for different effects that occur on comparable time scales as the translational diffusion. 
Consequently, the separation of time scales and the accurate determination of diffusion constants are impossible without specially engineered tools. 
This task is undertaken in the following section.

\begin{figure}
\includegraphics[width=\columnwidth]{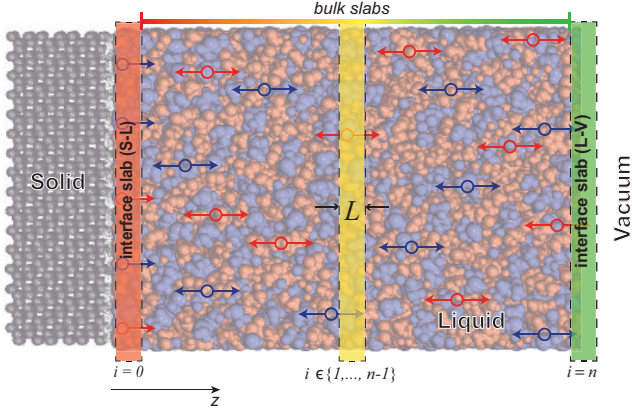}
\caption{\label{fig:scheme_system}\textbf{Scheme of the nanoconfined simulation box (S-L-V) containing $1800$ ion pairs:} The highlighted light red rectangle (left) represents a thin slab near the solid interface, light yellow (middle) corresponds to bulk-like slabs and light green (right) represents a thin slab near the vacuum interface. See text for details. \textit{(Figure adapted from \cite{vuvcemilovic2021computational})}}
\end{figure}

\section{Extensive Particle Model (EPM) for particles with internal degrees of freedom}

To rectify the discussed shortcoming of the SPM and similar approaches, we extend the previous theory \cite{spm_pub} to take into account more complex, structurally flexible molecules. 
This extension is kept very simple to be as general as possible and will be referred to as the \emph{Extensive Particle Model (EPM)} (see appendix \ref{app:theory_epm} for details). 

We start our analysis on a reduction of the liquid dynamics to movement along only one major axis, which we will refer to as the $z$-direction. 
The other dimensions are reduced under the assumptions of sufficient symmetry. 
In the presence of an interface (see e.g. \cref{fig:scheme_system}), we assume $z$ to be interface-orthogonal.
This effectively 1D-system can then be cut into smaller slices of thickness $L$. 
We will refer to the mean time that a particle starting within the slab continuously remains within the slice as \emph{lifetime} or $\tau$.
The goal is to determine the mean local diffusion coefficient $\langle D_\perp \rangle$ within a slice based on observed particle lifetimes within the slice. 
More complex extensions than a 1D-system are possible but beyond the scope of this work. 

To account for molecular flexibility and to establish a link between $\tau_{\mathrm{EPM}}$ and $\langle D_\perp \rangle$, we introduce a second coordinate $s$. 
This new coordinate denotes the offset from the position $z$, such that the particle is observed at $z+s$ due to its additional modes of transport, either induced by structural fluctuations or by coordinated interactions with the environment. 
Considering that particles, as well as particle clusters, are of finite size, we limit $s$ by a maximum absolute amplitude of the displacement $d_{\mathrm{mol}}$. 
The precise value of $d_{\mathrm{mol}}$ depends on the system and particle type. 
The time-dependent probability distribution $p$ on the space of configurations as described in \cite{spm_pub} is then extended to feature the new diffusive coordinate $s$ as $p(z,s,t)$.

\begin{table}[tb]
\caption{\label{tab:epm_considerations}%
\textbf{Model assumptions for the Extensive Particle Model (EPM):} Extensive Particle Model domains, parameters and assumptions for the bulk and LV interface slabs respectively. (BC: Boundary condition, $R_\mathrm{lim}$: lower bound for $R$)
}
\begin{tabular}{cc||c|c}
\toprule
\multicolumn{2}{c||}{Slab type} & Bulk (B) & Liquid-Vacuum (LV)\\
\midrule
\midrule
\multicolumn{2}{c||}{$D_{\perp}(z)$}	& $const$ & $const$\\
\midrule
\multicolumn{2}{c||}{Domain}	& \includegraphics[width=.3\columnwidth]{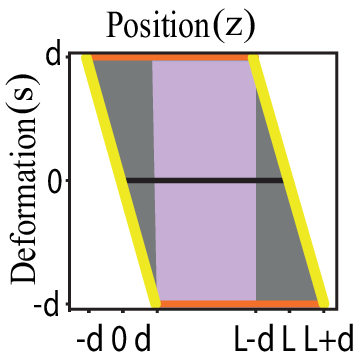} &  \includegraphics[width=.3\columnwidth]{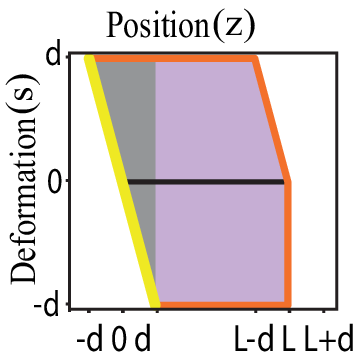}\\
\midrule
\multirow[c]{2}{*}{$z$-BC} & $z_i$ & absorbing & absorbing\\
& $z_i+L$ & absorbing & reflective\\
\midrule
$s$-BC & $\pm d_\mathrm{mol}$	& reflective & reflective\\
\midrule
\midrule
$\nu$& 	& \multicolumn{2}{c}{$D_{\mathrm{mol}}/D_\perp$}\\
\midrule
$q$& 	& \multicolumn{2}{c}{$d_\mathrm{mol}/L$}\\
\midrule
\midrule
\multicolumn{2}{c||}{$R_{\mathrm{lim}}$}	& \multicolumn{2}{c}{$\left[\max\left(1-2 q, 0\right)\right]^3$}\\
\bottomrule
\end{tabular}
\end{table}
\begin{figure}
    \centering
    \includegraphics[width=.48\textwidth]{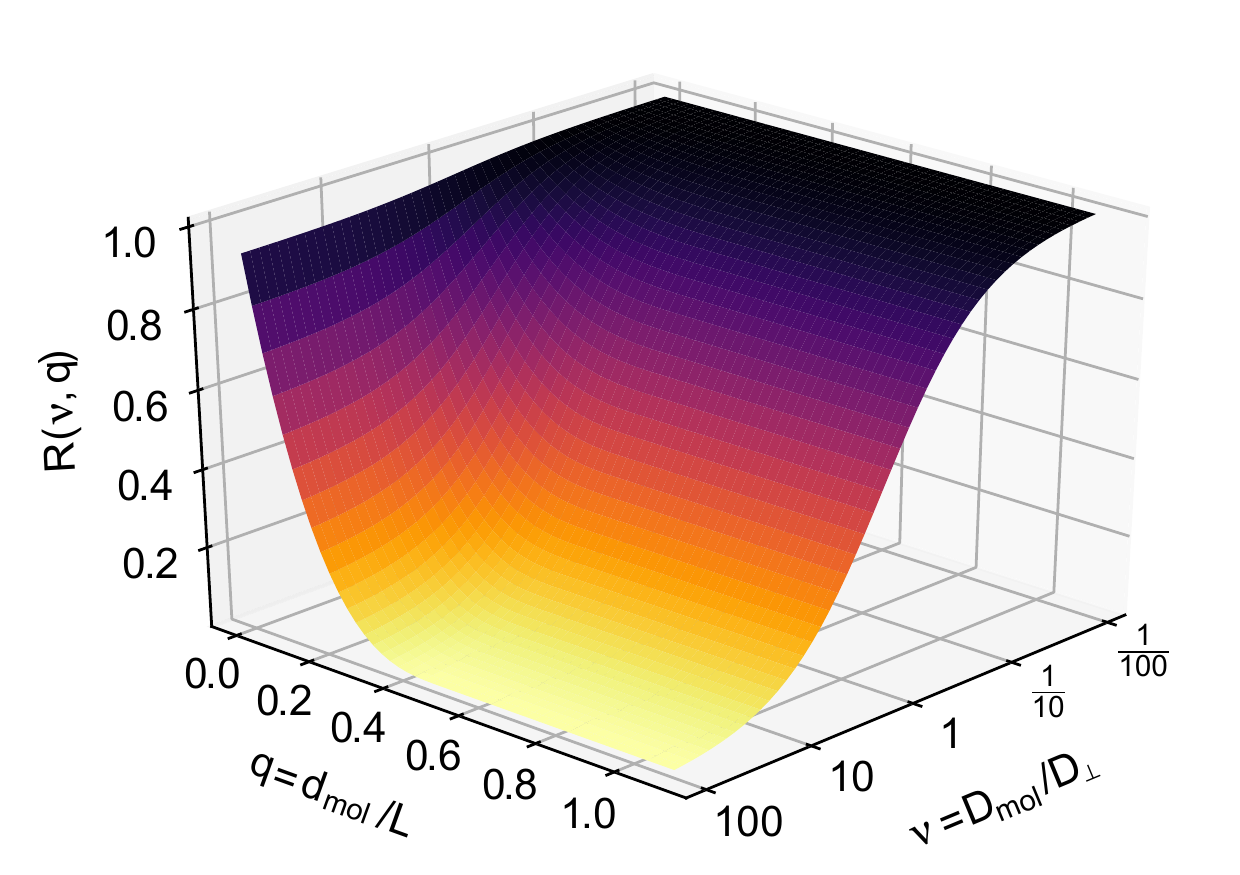}
    \caption{\textbf{The universal correction factor $R(\nu,q)$:} 
    Correction factor $R(\nu,q)$ obtained from numerical solutions of \cref{eq:extensive_fpe}, plotted for values $\nu\in[0.01,100]$, $q\in[0,1.1]$. 
    Cross-sections along the $\nu$- and $q$-axis respectively are provided in \cref{fig:app_correction_crossection}}
    \label{fig:universal_correction_function}
\end{figure}

\begin{figure*}
\includegraphics[width=2\columnwidth]{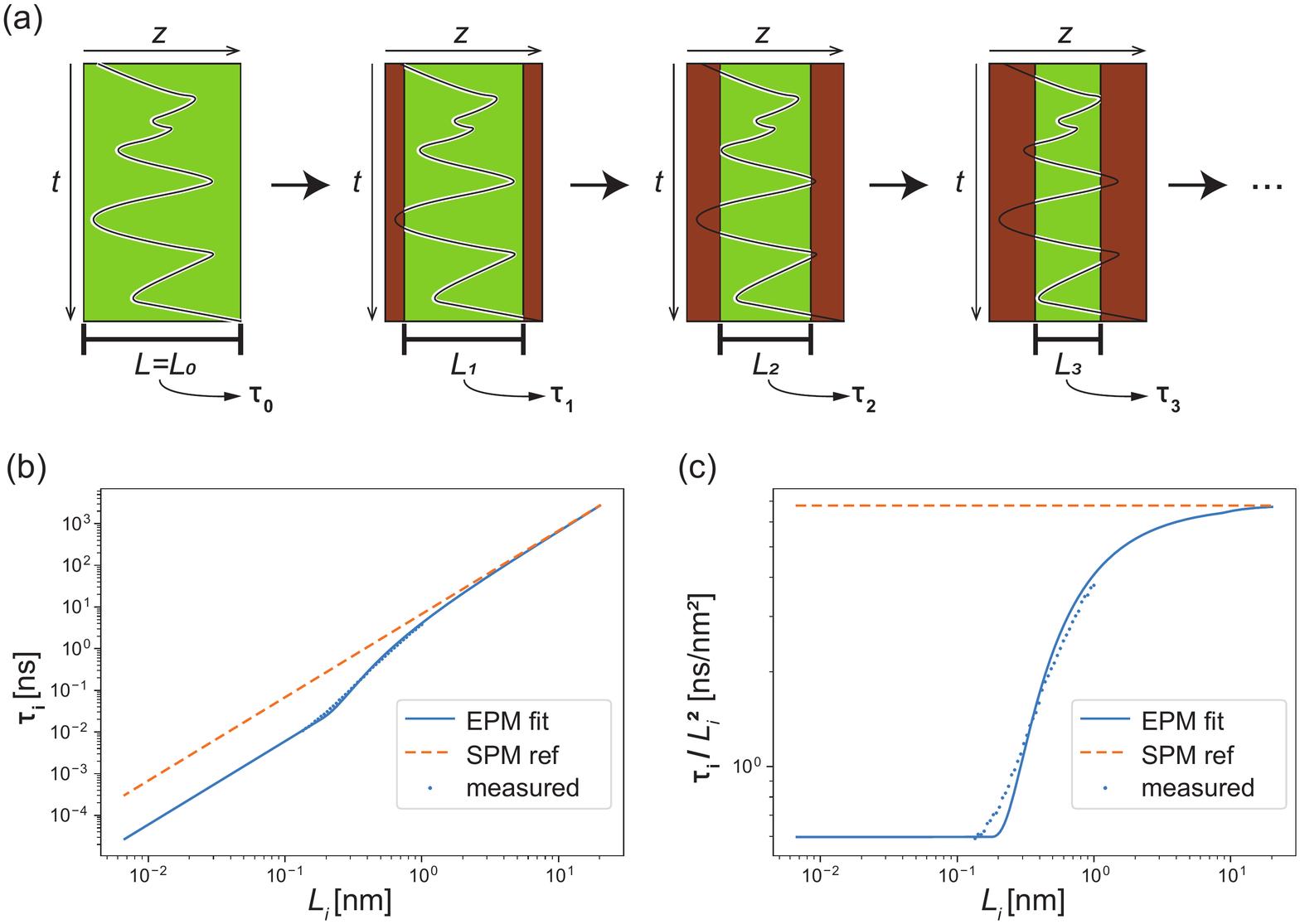}
\caption{\label{fig:width_reduction_epm_fit}
\textbf{The \emph{Local Width Reduction} (LWR) method and its applications:} 
(a) Visualisation of the LWR method to obtain $D_\perp$ within a slab of thickness $L=L_0$ without prior knowledge of the deformation parameters $D_{\mathrm{mol}}$ and $d_{\mathrm{mol}}$ by measuring mean lifetimes $\tau_i$ for successively thinner and thinner slices of thicknesses $L_i\leq L$. 
(b) Lifetime according to SPM for a particle with same diffusion coefficient $D$ (orange line, dashed), measured lifetimes $\tau_i$ (blue dots) and the EPM fit (blue line, solid) and (c) rescaled plot of $\tau$ normalised by the expected slice thickness dependence.}
\end{figure*}

In the EPM, a particle's state is therefore represented by 
$$(z,s)\in [z_i-d_{\mathrm{mol}},z_i+L+d_{\mathrm{mol}}]\times [-d_{\mathrm{mol}}, +d_{\mathrm{mol}}]$$
with the particle being observed at $z'=z+s$. 
As the position $z'$ now needs to be located within the slice instead of $z$, the precise restriction on allowed and considered configurations amounts to $z'\in [z_i, z_i+L]$. 

For a bulk-like slab, where the particle can exit on both sides, the domain takes a parallelogram-shape. 
Absorbing boundary conditions are applied at the tilted $z$-direction boundaries with reflecting boundary conditions in $s$-direction (see \cref{tab:epm_considerations}). 
For a slab at an interface with the solid or vacuum (see also \cite{spm_pub}), the particle cannot be at a $z$-position behind the interface, so the parallelogram-shaped domain gets one corner cut off, and the reflecting boundary conditions are extended to the boundary towards the interface, as visualised in \cref{tab:epm_considerations}.

To be able to model the dynamics of the displacement $s$, we assume that the motion in $s$-direction can be described via a diffusive and drift-free process, only introducing an additional \emph{deformation diffusion coefficient} $D_{\mathrm{mol}}$. 
We furthermore assume that $D_{\mathrm{mol}}$ depends on the particle type and the position of the slab but is constant across an individual slab. 
Consistently with the basic assumptions of the SPM \cite{spm_pub}, we again choose a uniform density distribution across the phase-space domain as the initial condition of our system. 
We then arrive at the following PDE:
\begin{equation}
\partial_t p (z,s,t)=\partial_z (D_\perp  (z) \partial_z p(z,s,t))+D_{\mathrm{mol}}  \partial_s^2 (p(z,s,t)),	\label{eq:extensive_fpe}
\end{equation}
which we solve up until the long time limit. 
At this point only one last mode of the eigenfunctions with exponential decay remains, which we then extrapolate.

The tilted shape of the phase-space, and its boundary conditions make a general analytical solution to this PDE inaccessible, but, based on a scaling argument, the degree to which $\tau_{\mathrm{EPM}}$ changes compared to $\tau_{\mathrm{SPM}}$ should only depend on the relative scales of disturbances, i.e.
$$\nu=\frac{D_{\mathrm{mol}}}{\langle D_{\perp}\rangle} \text{ and } q = \frac{d_{\mathrm{mol}}}{L}.$$ 
Here the diffusion ratio $\nu$ gives a notion of the rapidity of the molecular degrees of freedom compared to molecular translations, while the length ratio $q$ expresses the amplitude of the motion of the molecular centre of mass due to internal fluctuations relative to the slice size. 

We can thus write $\tau_\mathrm{EPM}$ analogously to \cref{eq:D_relation_basic_shape} with an additional term $R(\nu,q)$ denoting the relative change of lifetimes:
\begin{align}
    \tau_\mathrm{EPM} = R\left(\nu,q\right)\times c_\mathrm{slab} \frac{L^2}{\langle D_\perp \rangle}.	\label{eq:epm_solution}
\end{align}
Here $c_\mathrm{slab}$ again depends on the type of slice that is being investigated (i.e. $c_\mathrm{LV} = 1/3$ and $c_\mathrm{B}=1/12$), and $R(\nu,q)$ is the correction factor, which can be cast into a universal form, as discussed below.

\subsection{The correction factor $R(\nu,q)$}
The correction factor can be obtained by numerically solving \cref{eq:extensive_fpe} for a fixed choice of $\nu$ and $q$ (e.g. by setting $D_\perp=1$, $L=1$ and $D_\mathrm{mol}=\nu$ and $d_\mathrm{mol}=q$).
Specifically, we first calculate $\tau_\mathrm{EPM}(\nu,q)$ by solving \cref{eq:extensive_fpe} with appropriate boundary conditions. 
We then divide the result of \cref{eq:epm_solution} by the prediction of \cref{eq:D_relation_basic_shape} which yields $R(\nu,q)$ (see \cref{fig:universal_correction_function}). 
Notably, this presented correction factor $R(\nu,q)$ is universal, i.e. it does not have to be recalculated for different system setups but only once for each relative scale of disturbances. 

Since $L$ is determined by the slicing, it is sufficient to determine $D_\mathrm{mol}$ and $d_\mathrm{mol}$ in any system to be able to derive $D_\perp$ from \cref{eq:epm_solution}.
More specifically, With all parameters but $D_\perp$ in \cref{eq:epm_solution} being known, we effectively fix $q$ as a parameter of $R$ and only leave $\nu$ to vary, exploring a slice $R(\nu)_q$ of the universal correction function $R(\nu,q)$ (consult the github project page [\url{https://github.com/puls-group/diffusion_in_slit_pores}], the associated Zenodo archive \cite{hollring_kevin_2022_7446071} and the notes in appendix \ref{app:provided_software} for tabular data on $R_\mathrm{B}$ and a script to calculate further values).
The problem then boils down to finding the right value of $D_\perp$, such that the right hand side of \cref{eq:epm_solution} (i.e. $const \times R(\nu)_q/\langle D_\perp\rangle$) coincides with the mean lifetime $\tau$ of a molecule in that slab, measured in simulations or experiments. 

The obtained diffusion coefficient should not differ significantly from the results of the SPM as long as the amplitude of the relative centre of mass displacement is small ($q\to 0$). 
Likewise, no effect should be observed if the motion of the centre of mass due to structural fluctuations occurs at time scales that are negligible compared to the time scale for the translation alone ($\nu\to 0$). 
Hence:
$$R(x\to0,q)=R(x,q\to 0) = 1 \hspace{10pt}\text{(see \cref{fig:universal_correction_function})},$$
which is typically assumed by existing approaches as well as the SPM \cite{spm_pub}.

In the limit of fast structural fluctuations ($\nu\gg 1$), the time scales decouple. 
Any displacement in $s$ can be considered instantaneous on the time scale of diffusion in $z$. 
The molecular deformations can cause particles in certain states to leave the slab (gray regions in domains in \cref{tab:epm_considerations}), effectively reducing the slab width. 
As a result, we are thus left with the following lower bound for the high-$\nu$-limit (see appendix \ref{app:universal_R} for the derivation):
$$R_B(\nu\gg 1,q \leq 0.5) \geq (1-2q)^3.$$
At $q\geq 0.5$, the particle's centre of mass can and will be driven out of the entire slab by deformation on a faster time scale than via diffusion in $z$ (the shapes in \cref{tab:epm_considerations} become completely grey). 
In this case, we can only resolve $D_\mathrm{mol}$ and not $D_\perp$ as the dominant motion. 
The asymptotic behaviour thus changes to:
$$R_B(\nu\gg 1,q \geq 0.5) \propto \mathcal{O}(\nu^{-1}).$$

As mentioned previously, it is a simple task to resolve $D_\perp$ as long as $D_\mathrm{mol}$ and $d_\mathrm{mol}$ are known.
However, it is often a challenge to determine $D_\mathrm{mol}$ and $d_\mathrm{mol}$ in a molecule with many degrees of freedom all interacting with the environment. 
An additional issue is the coupling between translations and rotations.

\subsection{Systematic coarsening by slice width reduction} 

To circumvent the problem of determining  $\langle D_\perp\rangle$ at finite but unknown $D_\mathrm{mol}$ and $d_\mathrm{mol}$, we resort to gradual coarsening of the structural fluctuations by increasing their relative significance, in a process that we refer to as \emph{Local Width Reduction} (LWR). 
Specifically we continuously change $q$, while keeping $\nu$ fixed by systematically reducing the slab width as displayed in \cref{fig:width_reduction_epm_fit}a. 
We start from a large initial $L=L_0$ centred at $z=z_c$. 
The width $L_0$ should be at least comparable to $d_\mathrm{mol}$, but small enough such that $D_\mathrm{mol}$, $d_\mathrm{mol}$ and $D_\perp$ are still reasonably constant within the slab.
We then determine the sequence of the life times $\tau_\mathrm{EPM}(L_i)$ for smaller and smaller slabs centered at $z_c$ (i.e. $[z_c - L_i/2,z_c + L_i/2]$, with $L_{i+1}<L_i\leq L_0, \forall i\geq 0$, see \cref{fig:width_reduction_epm_fit} b).
In doing so we continuously change the scale $d_\mathrm{mol}$ of displacement relative to the scale of the slab $L$ in the measured sequence of lifetimes. 
Finally, the $\tau_\mathrm{EPM}(L_i)$ are used to calculate the deviations from the scaling predicted by the SPM, i.e. $$\tau_\mathrm{EPM}(L_i) \to \frac{\tau_\mathrm{EPM}(L_i)}{L_i^2}$$ 
(see \cref{fig:width_reduction_epm_fit}c) as this deviation is mainly determined by $R$.
In doing so, we generally only need to apply the method to bulk-like slabs, where $i\geq 1$, as only in the initial slicing of thickness $L_0$ the slice can be immediately adjacent to an interface, therefore only requiring the calculation of $R_B$. 

The set of 
$$\frac{\tau_\mathrm{EPM}(L_i)}{L_i^2}$$
is then fitted with the universal curve for $R(\nu,q)$ (\cref{fig:universal_correction_function}) with $D_{\perp}(z_c)$ as well as $d_\mathrm{mol}(z_c)$ and $D_\mathrm{mol}(z_c)$ as fitting parameters. 
Determining the appropriate line in the space of $R(q)|_\nu$ with only $q$ as a free parameter among the $\tau_\mathrm{EPM}(L_i)$, now sets $\nu$, and eventually $D_\mathrm{mol}(z_c)$, $d_\mathrm{mol}(z_c)$ thus fixing $D_{\perp}(z_c)$ (see appendix \ref{app:lwr_fit} for details on the chosen fitting procedure).

Performing the Local Width Reduction (LWR) scheme in different slabs throughout the system not only provides us with the spatially resolved diffusion coefficient for translation but also gives us insights into the evolution of molecular deformation processes as a function of the distance from a confining wall. 

\begin{table}[tb]
\caption{\label{tab:correction_model_benchmark_epm}%
\textbf{Bulk IL self-diffusion coefficients accounting for extensive particles:}
Here, $n$ denotes the number of carbon atoms in the alkyl chain of the cation.  
MSD data as from \cref{tab:correction_model_benchmark_spm} are compared to data from the EPM fit results obtained via LWR, using slices of initial thickness of $L_0=\SI{1}{\nano\meter}$. 
The error $\mathrm{EPM}_\mathrm{Err}$ is the deviation of the EPM relative to the MSD as in \cref{eq:msd_err}. 
All diffusion constants are in units of $\SI{e-7}{\centi\meter\squared\per\second}$.
}

\begin{tabular}{c|ccc}
\toprule
&\multicolumn{3}{c}{$D^+$}\\
$n$	& MSD & EPM &EPM$_{\mathrm{Err}}$ \\
\midrule
2  	& 
$3.5\pm 0.1$ & $3.5\pm 0.1$ & $\SI{0}{\percent}$\\
4	& 
$2.6\pm 0.1$  & $2.8\pm 0.1$ & $\SI{7}{\percent}$\\
6	& 
$1.4\pm 0.1$ & $1.6 \pm 0.1$ & $\SI{14}{\percent}$\\
8	& 
$0.8\pm 0.1$ & $1.2 \pm 0.1$ & $\SI{50}{\percent}$ \\
10	& 
$0.7\pm 0.1$ & $1.0 \pm 0.1$ & $\SI{40}{\percent}$ \\
\midrule
&\multicolumn{3}{c}{$D^-$}\\
$n$	& MSD & EPM & EPM$_{\mathrm{Err}}$\\
\midrule
2  	& 
$2.0\pm 0.1$ & $2.1\pm 0.1$ & $\SI{5}{\percent}$\\
4	& 
$1.7 \pm 0.1$ & $2.0\pm 0.2$ & $\SI{17}{\percent}$\\
6	& 
$1.1 \pm 0.1$ & $1.3 \pm 0.1$ & $\SI{18}{\percent}$ \\
8	& 
$0.7 \pm 0.1$ & $1.0 \pm 0.1$ & $\SI{43}{\percent}$\\
10	& 
$0.6 \pm 0.1$ & $0.9 \pm 0.1$ & $\SI{50}{\percent}$ \\
\bottomrule
\end{tabular}
\end{table}
Depending on the chosen parameters for a simulation, there are lower and upper limits to the values of $\tau$ that we can reliably resolve from simulation data.
Within this window of possible values, there need to be sufficiently many data points for the fit of $R$ to yield reliable results.
This needs to be taken into account for the setup of the simulation, but also the initial choice of the slice thickness $L_0$ as it will limit the possible spatial resolution of the diffusivity $D_\perp(z)$.
This resolution has, in our experience, proven to be significantly higher than what can be achieved by standard methods for the determination of spatially dependent transport coefficients, e.g. the Einstein method applied to parallel diffusion in a sliced system.

\section{Validation of the EPM by characterising diffusion in bulk ILs}
\label{spm_epm_validation}

We evaluate the performance of the EPM using molecular dynamics (MD) simulations of bulk imidazolium-based ionic liquids (IL), with MSD diffusivity data as seen in \cref{tab:correction_model_benchmark_spm} being used as a reference.
To calculate the predictions of the EPM, non-overlapping and adjacent slices are chosen with a fixed slice thickness $L$.
Here we specifically use the same thickness as employed for the SPM (\cref{tab:correction_model_benchmark_spm}), covering the entire simulation box.
The diffusion constant $D_\perp$ is calculated in each slice independently, then the values are averaged and the standard deviation is calculated across all slabs for comparison with $D_\mathrm{MSD}$.

Notably, the EPM results for $D_\perp$ become consistent with those obtained from the MSD  (\cref{tab:correction_model_benchmark_epm}). 
The error is less than $\SI{5}{\percent}$ for short chains and about $\SI{50}{\percent}$ for the largest cations.
The latter can be attributed to limitations of collapsing all internal degrees of freedom of long cations to a single degree of freedom modelled in the EPM.
The error is exacerbated by the decrease in the absolute values of the diffusion constant, making small absolute errors possibly induced by the numerical fitting method relatively more significant. 
This trend is also seen in the uncertainty of the diffusion constant calculated from the MSD, which also increased to $\SI{15}{\percent}$ for larger chains due to higher viscosity. 

\begin{table}
\caption{\label{tab:deformation_parameters_bulk_il}%
\textbf{Bulk IL self-diffusion coefficients and deformation parameters:}
Self-diffusion coefficients $D_\perp$ and deformation parameters $D_\mathrm{mol}$, $d_\mathrm{mol}$ from MD simulations for cations and anions (superscript $+/-$) obtained from EPM model fit results via Local Width Reduction analysis (LWR). 
The LWR results were extracted from slices of initial thickness of $L_0=\SI{1}{\nano\meter}$ in the pure bulk IL systems with subsequent averaging and error calculation across slices. (diffusion in units of $\SI{e-7}{\centi\meter\squared\per\second}$, $d_\mathrm{mol}$ in units of $\SI{e-2}{\nano\meter}$)
}
\begin{tabular}{c|ccc}
\toprule
&\multicolumn{3}{c}{Cation}\\
$n$	& $D^+_\perp$ & $D^+_\mathrm{mol}$ &$d^+_\mathrm{mol}$ \\
\midrule
2  	& 
$3.5\pm 0.1$ & $18.2\pm 0.1$ & $9.1\pm 0.1$\\
4	& 
$2.8\pm 0.1$ & $17.2\pm 0.1$ & $8.9\pm 0.1$\\
6	& 
$1.6\pm 0.1$ & $16.4\pm 0.1$ & $8.6\pm 0.1$\\
8	& 
$1.2\pm 0.1$ & $15.5\pm 0.1$ & $8.2\pm 0.1$ \\
10	& 
$1.0\pm 0.1$ & $14.7\pm 0.1$ & $8.6\pm 0.1$ \\
\midrule
&\multicolumn{3}{c}{Anion}\\
$n$	& $D^-_\perp$ & $D^-_\mathrm{mol}$ &$d^-_\mathrm{mol}$ \\
\midrule
2  	& 
$2.1\pm 0.1$ & $18.5\pm 0.1$ & $9.6\pm 0.1$ \\
4	& 
$2.0 \pm 0.2$ & $18.4\pm 0.1$ & $9.1\pm 0.1$\\
6	& 
$1.3 \pm 0.1$ & $17.4\pm 0.1$ & $9.0\pm 0.1$ \\
8	& 
$1.0 \pm 0.1$ & $17.6\pm 0.1$ & $9.1\pm 0.1$\\
10	& 
$0.9 \pm 0.1$ & $18.3\pm 0.1$ & $9.3\pm 0.1$ \\
\bottomrule
\end{tabular}
\end{table}

An added benefit of the EPM is its capacity to resolve $D_\mathrm{mol}$. 
Actually, several experimental studies pointed out the existence of two modes of diffusion \cite{singh2014ionic}.
A notable example comes from hf-BILs, a series comprised of bis(mandelato)borate anions and dialkylpyrrolidinium cations with long alkyl chains ($n=4-14$). 
In these systems, two diffusion constants were measured for ILs \cite{filippov2014effect}. 
The slower mode was similar to the diffusion of shorter alkyl chains, while a second an order of magnitude faster mode, sensitive to local viscosity, could be resolved but only for ILs with long chains ($n>10$). 
The result was associated with the appearance of the bi-continuous fluid phase. 
Our results suggest that this fast mode behaves consistently with $D_\mathrm{mol}$. 
As in these experiments, the here measured $D_\mathrm{mol}$ is sensitive to viscosity and decreases with increasing $n$ (see \cref{tab:deformation_parameters_bulk_il}). 
Notably, the ratio $\nu = D_\mathrm{mol}/D_{\perp}$ in our simulations increases from about $5$ at $n=2$ to over $20$ at $n=10$. 
This explains the reason for the observation of the fast mode only for ILs with long cations. 
At this point the fluctuations of the ion centre of the mass due to inter- and intramolecular interactions clearly occur on different time scales than the diffusive translations, and can distinctly be resolved. 
We show that the fast mode, still exists at low $n$, prior to the appearance of the bi-continuous phase. 
However, without explicit modelling as performed herein, it cannot be delineated from the slow translations.   

We conclude that the EPM proves generally capable of correcting for the limitations of the SPM. 
It provides meaningful absolute values of the translational diffusion coefficients, and captures the intricate dynamics of internal transport properties of complex liquids including ILs, as previously suggested in experiments. 
This establishes the EPM, in combination with the LWR, as a simple but essential method for understanding diffusive transport, well beyond the SPM and general Markov-State-Models used in the past.

\section{Anisotropic diffusion of ionic liquids in confined geometries}
Equipped with the proof that the EPM/LWR approach can resolve the self-diffusion coefficients of anions and cations in bulk systems of our considered ILs, we can now proceed to investigate the relation between ion structuring and the diffusive ion transport in IL thin films. 
This has been a challenging task experimentally and theoretically, because there is no clear separation of length scales between the molecular size of ions, the thickness of the film, and the effective interaction potentials between the solid phase and the liquid \cite{marion2021water}.

In the following, we want to illustrate the novel capabilities of the lifetime-based model for complex particles (EPM) in terms of resolving anisotropic diffusivities close to interfaces but also in the transitional region.
We will additionally discuss novel insights into the spatially dependent nature of internal degrees of freedom in the vicinity of interfaces.

\begin{figure*}[tb]
    \centering
    \includegraphics[width=.9\textwidth]{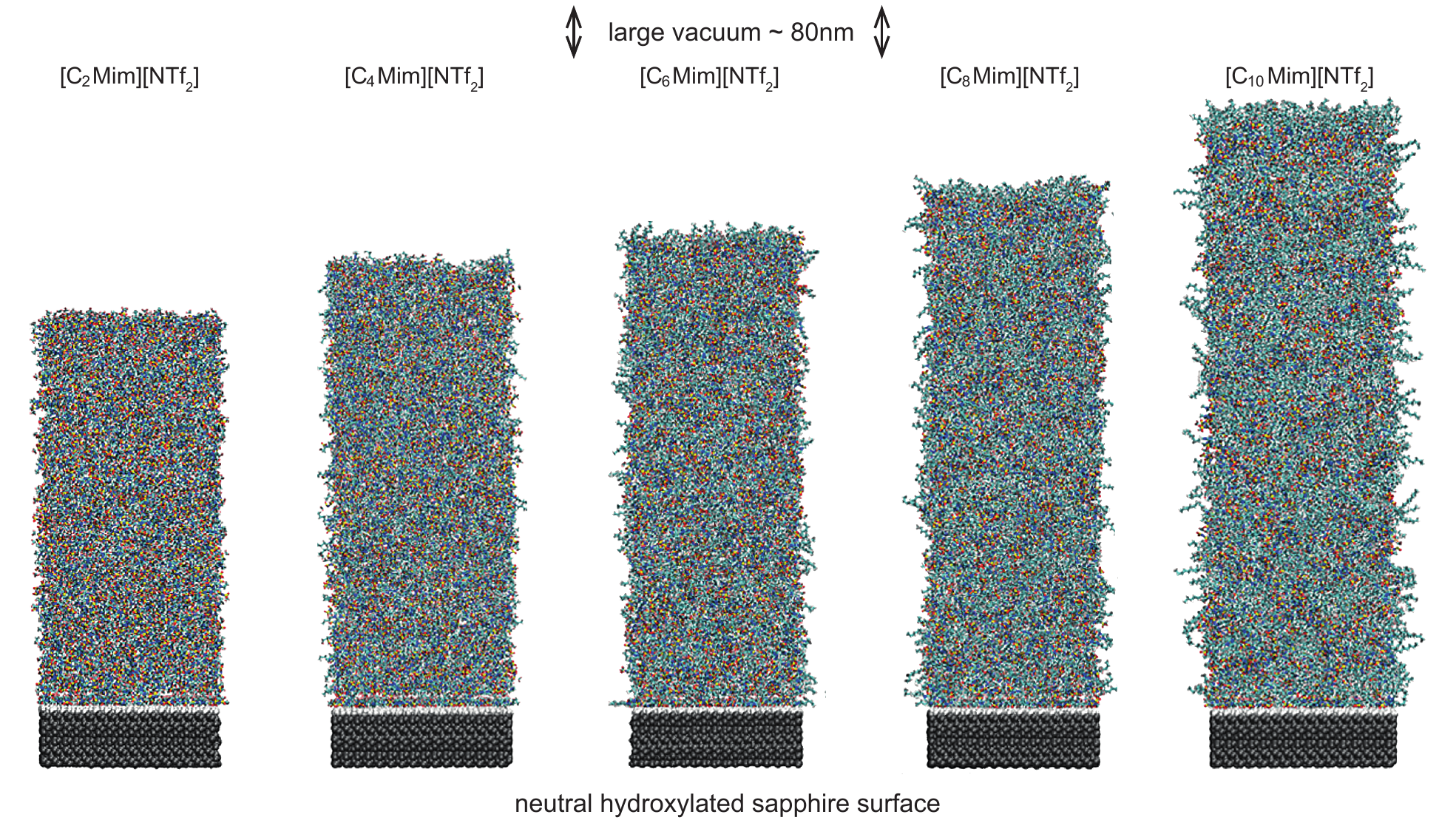}
    \caption{\textbf{The SLV system configuration:} Visualisation of the IL liquid films after equilibration on top of the hydroxylated sapphire surface below a $\SI{80}{\nano\meter}$ slab of vacuum for the different ILs [C$_n$Mim][NTf$_2$] ($n = 2,4,6,8,10$). 
    Clearly visible is the increasing film thickness with the increase in cation chain length. \textit{(Figure adapted from \cite{vuvcemilovic2021computational})}}
    \label{fig:app_slv_system_sim}
\end{figure*}

\subsection{MD simulations of thin IL films }
We first simulate thin IL films, following a previously established protocol \cite{vucemilovic2019insights}.
Accordingly, we position $1800$ [C$_n$mim][Ntf$_2$] ion pairs above a hydroxylated alumina slab in a monoclinic simulation box (see Appendix \ref{app:simulation_methods}). 
Since the alumina slab is identical in all systems, the thickness of the IL film increases with the size of the cation (\cref{fig:app_slv_system_sim}). 
To prevent $z$-replicas' effects on the formation of interfaces, a \SI{80}{\nano\meter} thick vacuum slab is placed above the liquid-vacuum interface. 
Following the annealing protocol in which the solid-liquid (SL), and the liquid-vacuum (LV) interfaces are fully formed, a \SI{200}{\nano\second} run is performed and used to build diffusion profiles (see appendix \ref{app:simulation_methods} for details).

\begin{figure*}[tb]
    \centering
    \includegraphics[width=0.95\textwidth]{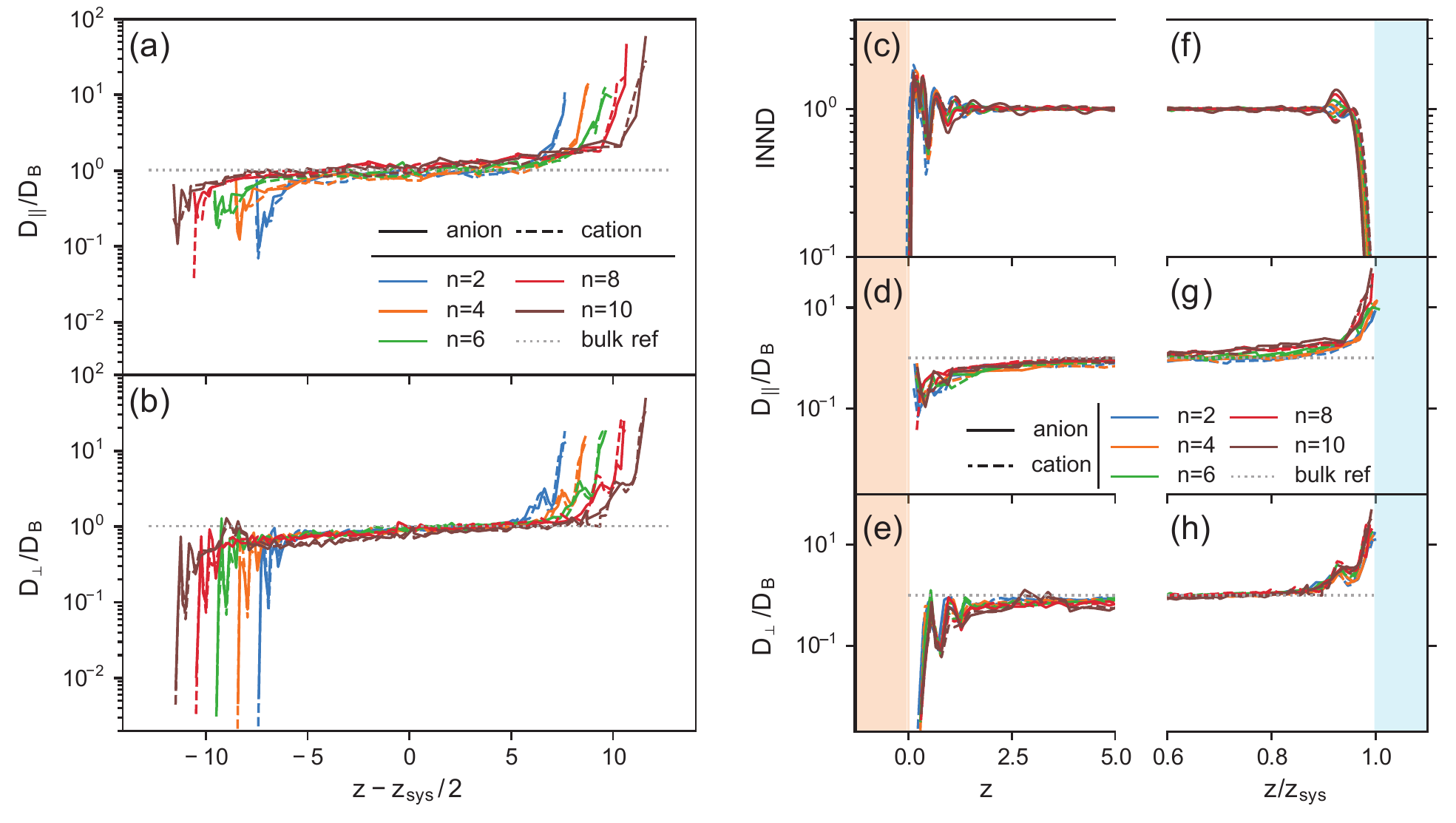}%
    \caption{\textbf{Particle transport in an S-L-V ionic liquid film:} (a) Interface-parallel $D_\parallel$ and  (b) -perpendicular self-diffusion $D_\perp$ for the IL SLV systems with the centre of the liquid film of thickness $z_{sys}$ shifted to zero. 
    (c,d,e) Plots of the normalised interface normal number density (INND) (c), normalised parallel diffusion $D_\parallel$ (d) and perpendicular diffusion $D_\perp$ (e) unscaled in $z$ direction, zoomed in on the SL-interface located at $z=0$.
    (f,g,h) Plots of the normalised INND (f), normalised $D_\parallel$ (g) and normalised $D_\perp$ (h) scaled with the film thickness $z_\mathrm{sys}$ in $z$ direction, zoomed in on the LV-interface, that is located at $z/z_\mathrm{sys}=1.0$ respectively.
    In all plots, dashed lines refer to cations, solid lines to anions and dotted to reference bulk values. 
    Different systems ([C$_n$Mim][NTf$_2$] for $n=2,4,6,8,10$) are distinguished via their colours . 
    $D_\parallel$ is normalised with respective bulk system MSD value, $D_\perp$ with the respective bulk system value obtained via EPM/LWR (see \cref{tab:correction_model_benchmark_epm}) and INND is normalised with the average bulk INND value.
    For the MSD analysis, $L$ is in the range $\SIrange{0.4}{1.0}{\nano\meter}$, for EPM/LWR the range of $L$ is $\SIrange{0.3}{1.0}{\nano\meter}$ with smaller values closer to the interface to resolve the features of the particle mobility there more accurately.
    } 
    \label{fig:IL_slv_innd_diffusion_profiles}
\end{figure*}

In focusing on the interface normal number density (INND) profiles of the ILs, we observe stable layering behaviour at the edges of the confinement/interfaces towards both solid and vacuum (\cref{fig:IL_slv_innd_diffusion_profiles} c and f, respectively). 
This matches well with AFM measurements of force-vs-separation profiles at these interfaces, where there is a characteristic dominant structure scale associated with each respective type of interface depending on local particle organisation \cite{ludwig2020recent}.
The higher order structure is also observable in the internal deformation parameters visualised in Appendix \cref{fig:app_deformation_IL_SLV}c and f, where we see the displacement amplitude $d_\mathrm{mol}$ drop towards either interface with a different characteristic range.
We therefore expect to observe different effects of these distinct dominant structures on the IL diffusivity close to the respective interfaces.

\subsection{Application of EPM}
We resolve the self-diffusion of the ILs with higher resolution at the interfaces compared to the bulk and connect our observations with already established understanding of the structuring properties of interfacial ILs. 

In the IL film systems, slices in the central bulk-like region are generally chosen to be $L=\SI{1}{\nano\meter}$ thick, where possible.
In the region close to the solid- and vacuum-interfaces respectively, smaller slices up to a lower limit of $L\geq \SI{0.3}{\nano\meter}$ for the perpendicular and $L\geq \SI{0.4}{\nano\meter}$ for the parallel direction were placed to again increase spatial resolution due to lower expected particle mobility with similar constraints as for the water system in \cite{spm_pub}. 
The spatially resolved diffusion profile parallel to the interface is obtained from the MSD, while the perpendicular component is obtained by applying the EPM in conjunction with LWR (\cref{fig:IL_slv_innd_diffusion_profiles} a and b, respectively). 
Both are normalised by diffusion constants as measured in the corresponding bulk liquids with the respective methods (see \cref{tab:correction_model_benchmark_epm}), to account for the systematic error of the $D_\mathrm{EPM}$.

Despite differences between systems, the $D$-profiles for anions and cations within the same system are almost identical. 
This was observed previously for interface-parallel diffusion for $n=2$ \cite{vucemilovic2019insights}. 
Such behaviour points towards the importance of the formation of ion pairs and other forms of IL internal structures due to the strong anion-cation Coulomb interactions \cite{iwata2007local,kirchner2015ion}. 

Throughout the system $D_\parallel$ normalised to the bulk value is consistently above the normalised $D_\perp$ both always exhibiting a slight yet persistent gradient throughout the central region of the film.
Interestingly the structure factor in that compartment is the same as in the bulk ILs \cite{vucemilovic2019insights}, however, as seen in confined water,\cite{spm_pub} dynamic properties seem to show relaxation on significantly longer length scales. 
The trends observed in the translational diffusion are systematically recovered by $D_\mathrm{mol}(z)$ (see Appendix \cref{fig:app_deformation_IL_SLV} for profile).  

Perpendicular components reach the reference bulk diffusion near the LV interface, while parallel components attain it closer to the solid interface. 
As a result, increased average parallel mobility of ions within the film is observed compared to the bulk IL. 
Such an effect of confinement has been reported previously, for example in phosphonium bis(salicylato)borate ILs in Vycor porous glass \cite{filippov2014effect},  or [C$_4$mim][Ntf$_2$] in mesoporous carbon \cite{chathoth2012fast} for both the slow and the fast mode of diffusion. 
Our result suggests that in contrast to the systems studied herein, filled pores will not show an increase in average diffusivitiy as also reported in our prior publication \cite{epm_pub}. 
Instead the increase in mobility relative to the bulk reference is associated with the presence of the vacuum interface.

\begin{figure*}[tb]
    \centering
    \includegraphics[width=.98\textwidth]{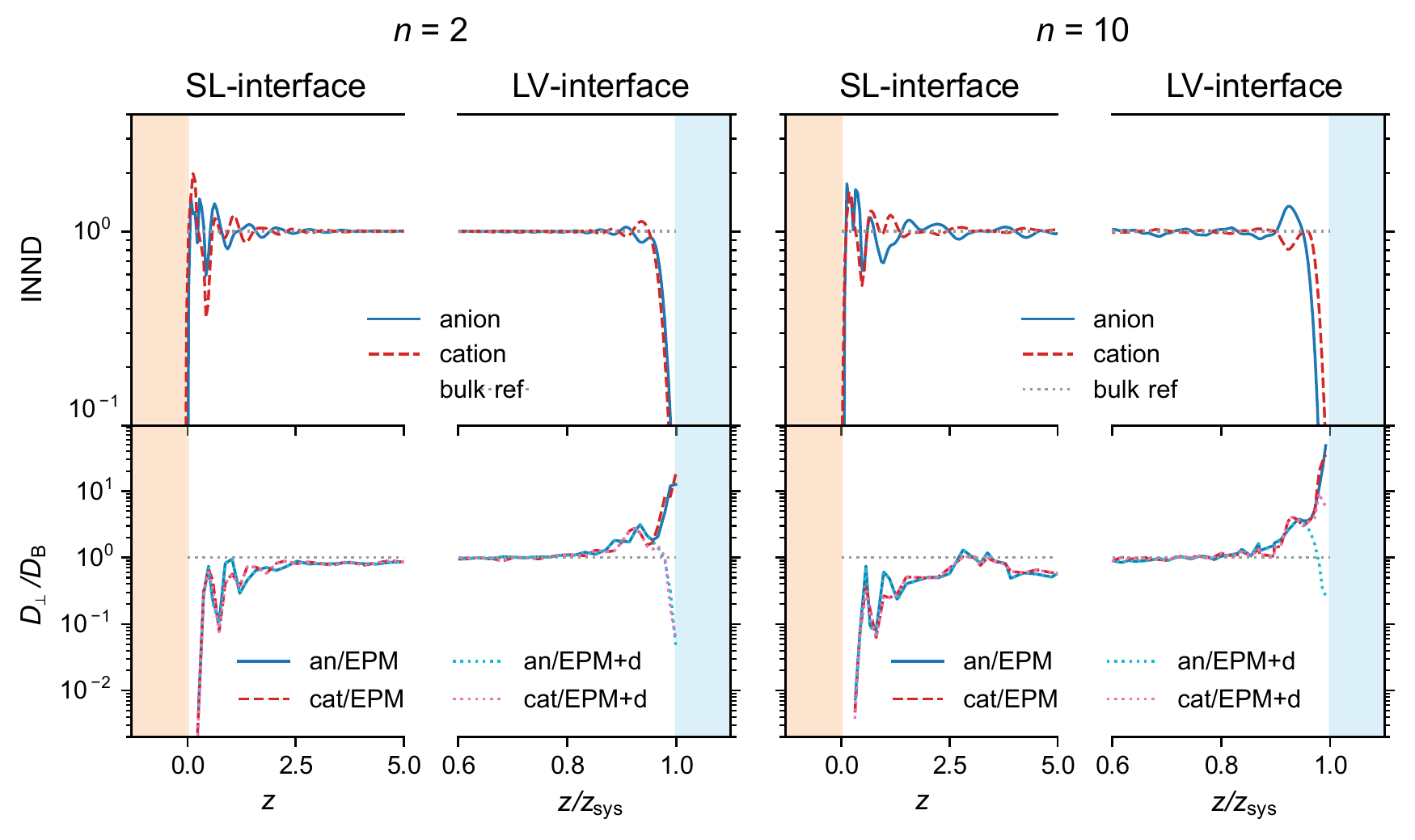}
    \caption{\textbf{Characteristic results of applying the drift-correction to $D_\perp$ of $n=2$ and $n=10$ systems:} (Left) Results for $n=2$, (Right) Results for $n=10$. 
    (Top row) INND of cations (red/dashed) and anions (blue/solid)  of ILs at the solid and vacuum interfaces relative to INND in bulk similar to \cref{fig:IL_slv_innd_diffusion_profiles}.
    (Bottom row) $D_\perp$ results according to the EPM for cations (red/dashed) and anions (blue/solid) relative to their bulk values as well as the results when applying the SPM+d correction factor $K(\gamma)$ (see main text for details). 
    Virtually no effect of the first-order drift correction can be observed at the SL-interface, whereas at the LV-interface, the ultimate increase in $D_\perp$ can be attributed to the drift of the steep effective potential at the vacuum interface. 
    The penultimate bump in diffusivity on the other hand cannot be attributed to the statistical drift, making it a significant observation.}
    \label{fig:app_correction_SLV_IL}
\end{figure*}

\subsection{Estimating the influence of drifts at interfaces}

Due to the potential of mean force acting between the IL and interfaces, which results in the INND, one can expect errors in the perpendicular diffusion coefficient because of the omission of potential gradients in the modelling of the EPM. 
Hence, it is important to discuss if the observed behaviour of locally varying diffusivity is significant or a consequence of a systematic error of ignoring the drift arising from the potential of mean force between the IL and the interfaces.

As derived in our previous publication \cite{spm_pub},
we can make an approximation of the combined effect of the extensive particle dynamics and the impact of drift, by multiplying the result of the EPM/LWR procedure by the correction term $K(\gamma)$ of the SPM+d approach (\cref{eq:D_spmd}) effectively resulting in an EPM+d approximation.
As described previously \cite{spm_pub}, the first order potential gradient can be extracted from the INND profiles. 
By applying this approximate first-order correction to the results presented in \cref{fig:IL_slv_innd_diffusion_profiles}, we conclude that, for all systems, the consistent oscillation patterns of particle mobilities at interfaces are a robust finding (see \cref{fig:app_correction_SLV_IL} for detailed results of $n=2$ and $n=10$). 
A similar result has previously been found for water confined to a slit pore in our precursor publication \cite{spm_pub}. 
We therefore arrive at the conclusion, that such oscillations appear for a wide range of simple to complex fluids 

\begin{figure}
    \centering
    \includegraphics[width=.45\textwidth]{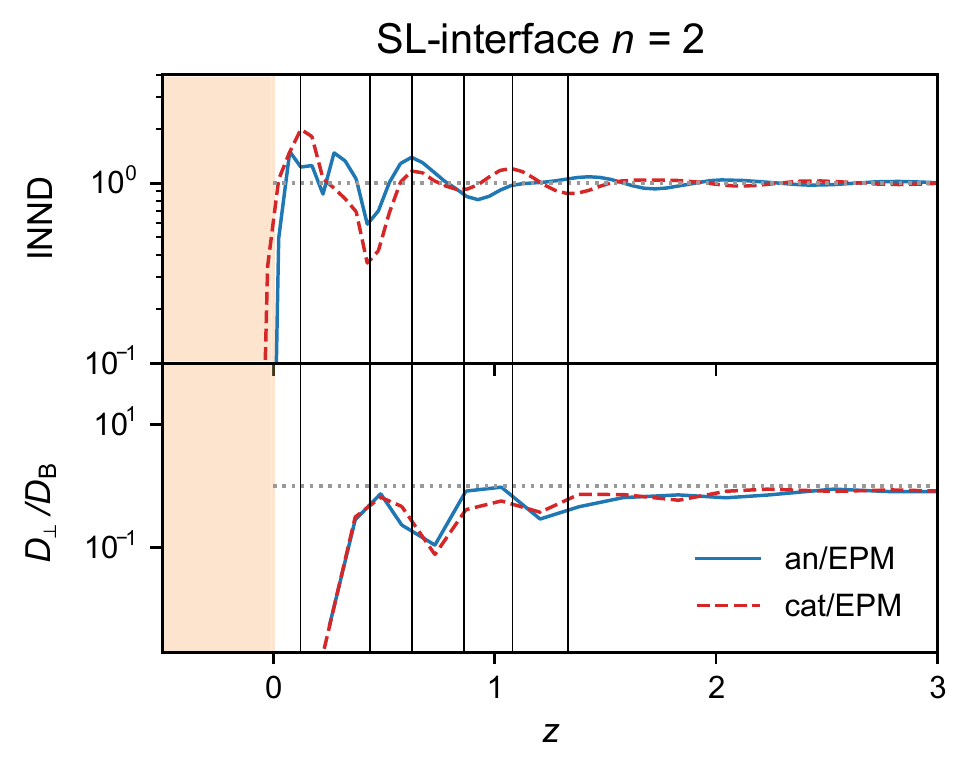}
    \caption{\textbf{Visualisation of the local extrema in INND and $D_\perp$ for the solid interface of $n=2$:} 
    Zoom on the SL-interface of $n=2$ in \cref{fig:IL_slv_innd_diffusion_profiles} to highlight the extrema in INND and $D_\perp$ not coinciding. 
    (Top) INND profile of cation (red/dashed) and anion (blue/solid) normalized by the bulk density.
    (Bottom) $D_\perp$ profile with the same color-coding based on the EPM results without drift correction due to little to no correction at the SL interface. 
    Vertical lines are centred on the local maxima/minima of the cation INND as a visual guide. 
    The extrema of the density profile do not all coincide with extrema of $D_\perp$, rendering the relation highly non-trivial.
    Similar observations are true for the other systems.
    }
    \label{fig:coincidence_c2_extrema}
\end{figure}

The comparison of the EPM result with and without the drift correction at the SL-interface does not show any significant change, meaning that the fluctuations clearly cannot be attributed to a first-order correction due to drift. 
Instead the observed oscillations are a direct consequence of the strict layering on top of solid interfaces. 
Importantly, while there is coupling of diffusivity to density oscillations, this effect is highly non-trivial, as the extrema of the two are not directly correlated (see \cref{fig:coincidence_c2_extrema}). 

The situation at the LV-interface is more delicate. 
Here, correcting for the drift shows that the ultimate increase of the diffusion coefficient $D_\perp$ in the very interface layer may be a spurious effect and that the diffusivity is overestimated by the EPM in that slab.
Still, the drift-correction leaves at least the penultimate mobility bump of $D_\perp$ as a significant result (see \cref{fig:app_correction_SLV_IL}).

\subsection{Diffusion at solid-liquid interfaces}

Equipped with the understanding that the EPM provides a good estimate for the diffusion profile at the SL interface, we can proceed to analyse interface-induced changes in the diffusion profile in more details. 
The effect of the SL interface is best observed in \cref{fig:IL_slv_innd_diffusion_profiles} (c,d,e), where the liquid density profile is shown together with both $D_\parallel$ and $D_\perp$ as a function of the distance from the pore wall. 
The position of the solid surface was fixed by the most extreme position of the centre of mass of an ion during the entire production run, all measured relative to the centre of mass of the entire film. 

All density profiles exhibit several characteristic oscillations as a function of the distance from the solid support as observed in other publications \cite{horn1988double,atkin2007structure}. 
Each layer corresponds to a peak in density (minimum in the effective IL-alumina interaction potential), while the dips of low density can be converted to free energy barriers for ions crossing between layers. 
While the absolute amplitude and the number of these oscillations is system dependent, their relative scale and frequency especially in the first layers is preserved as seen from the overlap in \cref{fig:IL_slv_innd_diffusion_profiles}c.  
As has been shown in our previous work \cite{vucemilovic2019insights}, the reason for this is the requirement of the electro-neutrality of each layer, the size of which is in principle set by the anion and the imidazolium rigid ring, common to all systems. 

\begin{figure}
    \centering
    \includegraphics[width=.45\textwidth]{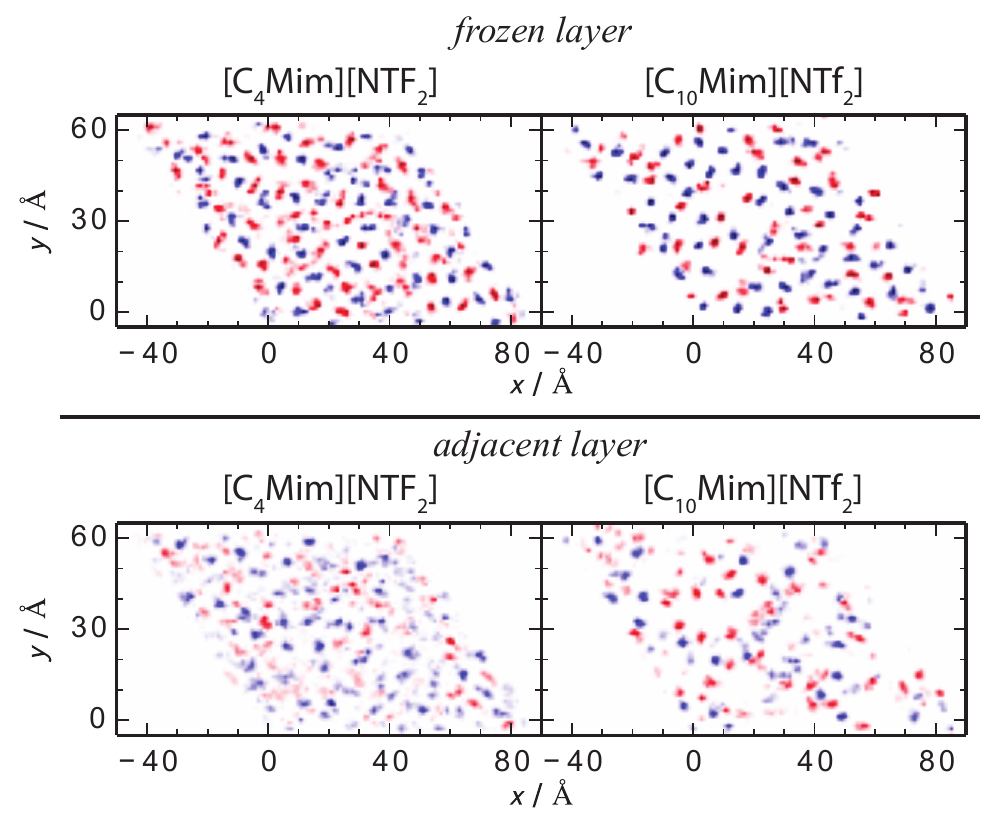}
    \caption{\textbf{Visualisation of the frozen layer on top of the solid for $n=4$ and $n=10$:} Residence probability distribution of anions (blue) and cations (red) in the frozen layer (first $\approx\SI{1}{\angstrom}$ on top of the SL interface) and the adjacent layer (next \SI{5}{\angstrom}) for $n=4$ and $n=10$. Visualised is the residence probability for the centre of mass of anions and cations in the layer at $t=0$ within a \SI{50}{\nano\second} time window. For both $n=4$ and $n=10$, ions remain mostly static within the observation time, whereas in the thicker adjacent layer, particles are more likely to move around and leave the layer. See appendix \cref{fig:frozen_layer_c2_c6_c8} for data on $n=2,6,8$ \textit{(Figure adapted from \cite{vuvcemilovic2021computational})} }
    \label{fig:frozen_layer_c4_c10}
\end{figure}

As seen before in water-filled pores \cite{spm_pub}, $D_\perp$ strongly anti-correlates with local normal density with its own oscillations superimposed over the overall drop-off towards the solid wall (\cref{fig:IL_slv_innd_diffusion_profiles}e). 
Within the barriers the particles are mostly observed crossing between adjacent layers and the $D_{\perp}$ is fast. 
Within the layers the stronger particle-particle interactions due to increased local density reduce ion mobility.

Due to a similar coupling with the local density $D_\parallel$ also experiences a reduction in mobility (\cref{fig:IL_slv_innd_diffusion_profiles}d). 
This reduction is less intense than $D_\perp$ because there is no confining potential for excursions along this direction of motion. 
However, the coupling with density is much stronger than in water, due to more significant inter-ion interactions. 
This overall reduction in particle mobility as well as the loss of mobility in internal degrees of freedom (see Appendix \cref{fig:app_deformation_IL_SLV}b and c) agrees with experimental findings on imidazolium based ILs at the solid-liquid interface in silica pores \cite{han2013distribution}.

Toward the very contact of IL with the alumina wall, ions make hydrogen bonds with the otherwise neutral solid \cite{segura2013adsorbed}, enabling the formation of an electro-neutral, nearly frozen layer of ILs.
This  attraction organises the ions into electro-neutral chequerboard patterns (see \cref{fig:frozen_layer_c4_c10}) \cite{brkljaca2015complementary}, from which the ions are unable to escape. 
Consequently, $D_{\perp}$ is reduced by more than two orders of magnitude compared to the reference bulk value, despite the strong drop in density. 
There is also a significant drop in $D_\parallel$, which however returns to the reference value at the very interface. 
This result probably stems from the lower resolution of the MSD at the interface compared to the EPM/LWR, which leads to the MSD averaging the very first layer as well as first inter-layer space therefore over-estimating particle mobility of interface ions. 

Actually, both, the parallel and the perpendicular component are highly suppressed, which can be seen from the tracks of the positions of ions' centres of mass over \SI{50}{\nano\second} (\cref{fig:frozen_layer_c4_c10}). 
Because of hydrogen bonds, ions basically move only sporadically, which is evident from individual non-overlapping islands for each ion at the contact with the interface. 
This is also the reason why this layer is not included in the EPM analysis (see \cref{fig:IL_slv_innd_diffusion_profiles,fig:app_correction_SLV_IL}). 
These results are consistent with the ultra-slow dynamics observed recently for $n=4$ at mica surfaces \cite{li2022extremely}, when, however, the attraction was provided by Coulomb interactions.

\subsection{IL diffusion at liquid-vacuum interfaces}

Moving on to the LV interface, we observe an entirely different behaviour. 
In \cref{fig:IL_slv_innd_diffusion_profiles} (f,g,h), the interface effects scale with the liquid film thickness $z_\mathrm{sys}$. 
This stems from the fact that the cations (dashed lines in \cref{fig:IL_slv_innd_diffusion_profiles} f) at the LV interface orient their non-polar alkyl chains towards the vacuum similar to other imidazolium-based ILs in MD studies \cite{lynden2006simulation}. 
This creates a shielding electro-neutral surface \cite{vucemilovic2019insights}.

Anions, on the other hand coordinate with the cation rings below the alkyl chains (full lines in the INND profiles (\cref{fig:IL_slv_innd_diffusion_profiles} f). 
Due to this orientation and alignment, the scale of the first molecular layer is determined by the length of the cations' alkyl chain. 
The latter also controls the thickness of the overall liquid phase at a constant ion pair count, hence the characteristic length scale for ordering at the LV interface, and the scaling of $D_\parallel$ (\cref{fig:IL_slv_innd_diffusion_profiles} g) and $D_\perp$ (\cref{fig:IL_slv_innd_diffusion_profiles} h)  in this region. 

The LV interface affects the diffusivities the most in the isolation region of the first molecular layer. 
Even after taking into account the drift (see \cref{fig:app_correction_SLV_IL}), both $D_\parallel$ and $D_\perp$ increase significantly compared to the bulk, while for $D_\parallel$ up to two orders of magnitude changes are observed.  
This effect gets stronger with the increasing chain length as the region with statistical absence of ions' charge centres thickens.

Different behaviours of $D_\parallel$ and $D_\perp$ are observed in the charged region of the first molecular layer, which coincides with the peak of liquid INND. 
Here, there is an up to \SI{10}{\percent} increase of overall liquid density despite an up to \SI{20}{\percent} lower cation density.  
In this region $D_\parallel$ changes slope and continues to drop towards the solid following a rather constant gradient that is surprisingly long-range. 

To the contrary, from the interface inward, after its drop-off towards the vacuum when accounting for the drift, $D_\perp$ first has a penultimate peak in all systems almost coinciding very closely with the maximum in total INND.
This peak is likely promoted by the anisotropic interactions as well as charge and density changes in the region, as symmetry breaking local structure in the vicinity of colloids in solution has also been proven experimentally to induce anisotropic Brownian diffusion contrary to an unstructured background \cite{king2006anisotropic}.
However, the overall effect of the LV interface on $D_\perp$ seems to be rather short-ranged as already in the second molecular layer, the ions experience the same particle-particle-interactions in perpendicular direction as elsewhere in the bulk liquid as also demonstrated by a quick recovery of the bulk deformation characteristics (see \cref{fig:app_deformation_IL_SLV}e and f).

\section{Discussion and Conclusions}

In this work, we introduce a novel technique for quantitatively analysing anisotropic diffusion in confined geometries, focusing on complex liquids, specifically ionic liquids (ILs). 
We demonstrate that internal molecular degrees of freedom can lead to additional spatially dependent modes of transport, comparable to self-diffusion, which have been overlooked in previous diffusion analyses. 
The latter include Markov-State-Model spectrum calculations \cite{prinz2011markov}, considered the most adaptable tool for studying transport in confined geometries \cite{voisinne2010quantifying}, as well the Simple Particle Model which we developed in the prequel to this paper \cite{spm_pub}. 

Our novel approach presented here overcomes the limitations of these models. 
Specifically, the here presented Extensive Particle Model (EPM) combined with Local Width Reduction (LWR) addresses the challenge of flexible molecules or particles deforming, fluctuating, and diffusing in confinement. 
Consequently, the EPM in conjunction with drift corrections represents a significant advancement compared to previous efforts in this field. 
It not only considers spatial variations in diffusivity but also accounts for the coupling of potentially spatially dependent degrees of freedom with translational transport processes, and underlying potentials. 
This improvement distinguishes the EPM from prior approaches that either neglect the impact of these coupled processes or assume strict separability \cite{bourg2011molecular,prinz2011markov}.

In applying the EPM to the complex transport processes in bulk ILs we have first proved its validity. 
Based on these results, we are now able provide an explanation for multiple different modes of transport previously detected in NMR experiments \cite{singh2014ionic,filippov2014effect}.
Specifically, we can explain why the ability to detect these modes has been dependent on the length of the cation chain of the IL.

After validating the EPM, we have applied it to thin films of imidazolium-based ILs. 
By exploring the correction that arises due to drifts induced by effective interaction potentials with the interfaces \cite{spm_pub}, we show that the EPM alone provides an excellent estimate of the diffusion profiles, particularly at the solid-liquid interface. 
The exception is the last molecular layer at the LV interface, where drift corrections may be significant.

We concede that our drift correction is a simplified approach applied directly to the translational degree of freedom, as derived in SPM \cite{spm_pub}.
However, this is precisely where we expect the majority of the effect. 
It is possible, however, to incorporate the drift into the  EMP starting from an adapted Smolouhowsky equation. 
However, proceeding with this calculation would affect the universal nature of the correction factor $R$, and significantly increase the complexity of application of the method.   
In the current version, a balance between the accuracy of the model and the reliability of its implementation is potentially achieved.

An obvious limitation of our here established approach similar to the SPM \cite{spm_pub} is its requirement for a subspace that can be reduced to a 1D representation for $D_\perp$ to be resolved. 
This is a restriction compared to other techniques, but it can be overcome by extending the model to a square or even a cube for possible grid-like analysis of diffusion. 
However, such an extension would require clearly establishing the role of coupling with different directional processes, which may be a challenge.

In using the EPM for confined ILs, we provide a deeper understanding of the coupling of the ions' translational diffusion with density stratification, the attractiveness of the interface, the spatial charge distribution, and additional internal degrees of freedom. 
Specifically, we clearly demonstrate a nontrivial link between the anisotropy in diffusivity and the anisotropic structuring of ILs at the contact with vacuum and the solid support.
We, furthermore, show different characteristic scaling behaviour of parallel and perpendicular diffusion coefficients at either of the two types of interfaces.
At the interface with the solid, the scaling is imposed by the electroneutrality of each layer and the cation ring diameter, while on the vacuum interface, the size of the hydrophobic cation chain dictates the length scale.
Finally, we find that interfaces may, but need not introduce effects in transport on a range significantly beyond the ones associated with density stratification, depending on the direction of transport and the type of interface. 
This result likely stems from hydrodynamic effects, and the effective stickiness of the surfaces in question \cite{baer2022modelling}.

To summarise our progress, this paper and its prequel \cite{spm_pub} establish universal tools that can be easily applied to study diffusive transport in a broad family of complex bulk and confined liquids. 
Using these tools, we can now deconvolve processes occurring on time scales similar to translation, and, in a quantitative manner, calculate the spatially resolved diffusion coefficients parallel and perpendicular to a pore wall. 
With necessary scripts publicly available, we hope these tools will find broad use in studies of molecular liquids.

\printcredits

\section{Acknowledgements}
The authors declare no competing financial interest.

We acknowledge funding by the Deutsche Forschungsgemeinschaft (DFG, German Research Foundation) – Project-ID 416229255 – SFB 1411 Particle Design and - Project-ID 431791331 - SFB 1452 Catalysis at Liquid Interfaces (CLINT).
The authors gratefully acknowledge the scientific support and HPC resources provided by the Erlangen National High Performance Computing Center (NHR@FAU) of the Friedrich-Alexander-Universit\"{a}t Erlangen-N\"{u}rnberg (FAU).

\appendix

\renewcommand\thefigure{A\arabic{figure}}%
\renewcommand\thetable{A\arabic{table}}%

%
%
%
%
%
%

\section{Extensive Particle Model } \label{app:theory_epm}

As the SPM  proved accurate for the simple water molecule  but insufficient for the more complex IL molecules, a model for corrections to the simple formula 
\begin{equation}
    D = \frac{1}{12}\frac{L^2}{\tau} \label{eq:default_connection}
\end{equation}
needed to be devised to account for the influence of particle deformation on the crossing time of particles within a subspace.
We can attribute this to the fact that an additional imprinted fast deformation movement of the centre of mass on top of the diffusive motion will lead to the particle leaving the subspace earlier than by diffusion alone, leading to a reduced expected lifetime $\tau'$. 
This reduced lifetime $\tau'$ incorporating the effects of deformation is the one that can be observed, whereas the formula \cref{eq:default_connection} expects the non-observable deformation free expected value $\tau$ to be measurable. 

It is therefore essential to find a link between $\tau'$ and $D$ along the lines of \cref{eq:default_connection} or -- as an approximation -- devise a link between $\tau$ and $\tau'$ such that we can calculate the virtual value of $\tau$ from the observed value $\tau'$.

To do this, we suggest a simplified model, where we have one variable $z$ describing the position of the reference point of the particle we are observing in interface-normal direction, as well as a second variable $s$ describing the displacement of the actually observed position of the particle relative to the reference point. 
The reference point $z$ is assumed to diffuse freely according to the drift-free Smoluchowski equation with the variable $s$ being confined in absolute value by an absolute maximum displacement $d$ defined by the geometry/structure of the particle. 
So we get $s\in[-d,d]$ as a restriction and need to detail how the displacement evolves over time. 
Again we choose a diffusive statistical model, where the displacement $s$ statistically performs a diffusive drift-free motion on the interval $[-d,d]$ with reflecting boundaries at $s=-d$ and $s=d$.
The diffusive constant $D_\mathrm{mol}$ in $s$-direction required for the Smoluchowski equation needs to be derived from the structure of the particles, ideally from the same simulation trajectory as the lifetime data.
We then arrive at a total PDE, describing the problem:
\begin{align}
    \partial_t \rho(z,s,t) = (D \partial_z^2+ D_\mathrm{mol} \partial_s^2) \rho(z,s,t) 
\end{align}
The relevant aspect here are the remaining boundary conditions and the shape of the domain on which the PDE needs to be solved.
As we observe the position of the particle at $z_\mathrm{tot} = z+s$ and we restrict $z_\mathrm{tot}\in[0,L]$, we observe the domain for an inner slab to be of parallelogram shape (see \cref{tab:epm_considerations}) determined by 
\begin{align}
I_d=\{(z,s)| z\in[-s,L-s], s\in[-d,d]\}
\end{align}
At an interface slab the option for $z<0$ is not possible due to the interface being present, so the domain at an interface will be given by
\begin{align}
I_d^\mathrm{int} = \{(z,s)| z\in[\max(0,-s),L-s], s\in[-d,d]\}
\end{align}
which takes the shape of a parallelogram with one corner cut off (see again \cref{tab:epm_considerations}).
The boundary condition for the inner slabs are assumed to be reflecting at $s=\pm d$ and absorbing at $z=-s$ and $z=L-s$.

For the boundary slab, the boundary at $z=L-s$ is assumed to be absorbing, whereas all others are considered reflecting as the particles cannot leave the slab there.

Then we need to again determine the survival probability $p(t)$ and calculate the mean survival lifetime $\tau$.
Unlike in the previous model attempt, this PDE cannot be solved analytically on our distorted domain, thus requiring solving via numerical methods starting from uniform initial probability distribution on the domain.

We can then compare the lifetime results with our prediction formula of according to the SPM \cite{spm_pub}, i.e. \cref{eq:D_relation_basic_shape}, to derive an approximate correction for a wide range of parameters.
Refer to the main manuscript for a discussion of the resulting correction factor $R(\nu,q)$.

%
%
%
%
%
%

\section{The universal correction factor $R(\nu,q)$} \label{app:universal_R}

\subsection{Bulk correction limit at $q\leq 0.5$}

In the case of bulk slabs as portrayed in \cref{tab:epm_considerations} we can make a prediction for the behaviour of $R(\nu,q)$ in the limit $\nu \to \infty$ for $q\leq 0.5$. 
Essentially, the fast speed of changes of configuration ($D_\mathrm{mol}$) compared to the spatial translation $D_\perp$, i.e. $\nu \gg 1$, causes the particles in the grey triangles of the domain visualised in \cref{tab:epm_considerations} to be immediately driven out of the slice by deformation alone. 
By this, we lose all particles that are not in the rectangular (pink) middle section of the slice. 
The ratio of these remaining particles is $$\frac{(L-2d_\mathrm{mol}) 2d_\mathrm{mol}}{L 2d_\mathrm{mol}} = 1-2q.$$
Only these remaining particles contribute to the non-zero mean particle lifetime. In addition, particles that diffuse into the grey area via $D_\perp$-related motion will also be driven out of the slab by conformation changes on a short/instant timescale. 
This effectively reduces the slab width to the $L-2d_\mathrm{mol}$ thickness of the centre region (pink) of the slab. 
The large value of $\nu$ also causes all statistics in $s$-direction of the domain to be equilibrated, so we can predict the mean lifetime for the particles in this region using the SPM approach.
The mean lifetime in this centre region is then 
$$\tau_\mathrm{SPM}(L-2d_\mathrm{mol}) = \frac{1}{12}\frac{(L-2d_\mathrm{mol})^2}{D_\perp}$$
compared to the translation-only lifetime of a bulk slab of thickness $L$: 
$$\tau_\mathrm{SPM}(L)=  \frac{1}{12}\frac{L^2}{D_\perp}$$
The mean lifetime in the slab $\tau_\mathrm{EPM}$ under these conditions can then be calculated as:
$$\tau_\mathrm{EPM} \geq \underbrace{2q\cdot 0}_{\text{grey area}}+ \underbrace{(1-2q)\cdot \tau_\mathrm{SPM}(L-2d_\mathrm{mol})}_{\text{pink area}}$$
therefore resulting in the following limit of $R$:
\begin{align}
    R(\nu \to\infty, q\leq 0.5) =& \frac{\tau_\mathrm{EPM}}{\tau_\mathrm{SPM}} \\
    \geq& (1-2q) \frac{(L-2d_\mathrm{mol})^2}{L^2} \\
    =& (1-2q)^3
\end{align}

\subsection{Approximations of $R_\mathrm{B}$ at $\nu\gg1$}

The limit $\nu\gg 1$ describes systems with dominant motion processes in terms of internal deformation on a shorter time scale than translation alone. 
We can then split the mean lifetime $\tau_\mathrm{EPM}$ into two contributions:
\begin{align}
    \tau_\mathrm{EPM} \approx& \tau_\mathrm{lim}(L,d_\mathrm{mol},D_\perp) + \tau_\mathrm{mol}(L,d_\mathrm{mol},D_\mathrm{mol})
\end{align}
with 
\begin{align}
    \tau_\mathrm{lim}(L,d_\mathrm{mol},D_\perp) =&\begin{cases}
        \frac{1}{12} \frac{(L-2d_\mathrm{mol})^3}{D_\perp L}& q\leq 0.5\\
        0&\text {else}
    \end{cases}
\end{align}
being the translation based lifetime only in the centre region where deformation cannot drive the particle out of the slice (pink, if it exists) and $\tau_\mathrm{mol}$ modelling the lifetime of a particle being driven out of the slice by deformation in the edge region (grey).
Due to the processes in the grey region being dominated by $D_\mathrm{mol}$, we can deduce that 
\begin{align}
    \tau_\mathrm{mol} = \mathcal{O}\left(\frac{1}{D_\mathrm{mol}}\right)
\end{align}
and hence 
\begin{align}
    R_B \approx \frac{\tau_\mathrm{EPM}}{\tau_\mathrm{SPM}} = R_{\mathrm{B},\mathrm{lim}}(q)+ \mathcal{O}(\nu^{-1})
\end{align}
where  
\begin{align}
     R_{\mathrm{B},\mathrm{lim}} = \begin{cases}
        (1-2q)^3& q\leq 0.5\\
        0&\text {else}
    \end{cases}
\end{align}
denotes the convergence limit for $\nu\to\infty$ for a fixed $q$. 

Due to this convergence behaviour, we have opted to only calculate $R(\nu,q)$ up to a maximum value of $\nu_\mathrm{max}=10^2$ and then extrapolate the values as follows:
\begin{align}
    R_B(\nu > \nu_\mathrm{max},q) = R_\mathrm{lim} + (R_B(\nu_\mathrm{max},q) - R_\mathrm{lim}) \frac{\nu_\mathrm{max}}{\nu}
\end{align}

\begin{figure}
    \centering
    \includegraphics[width=.45\textwidth]{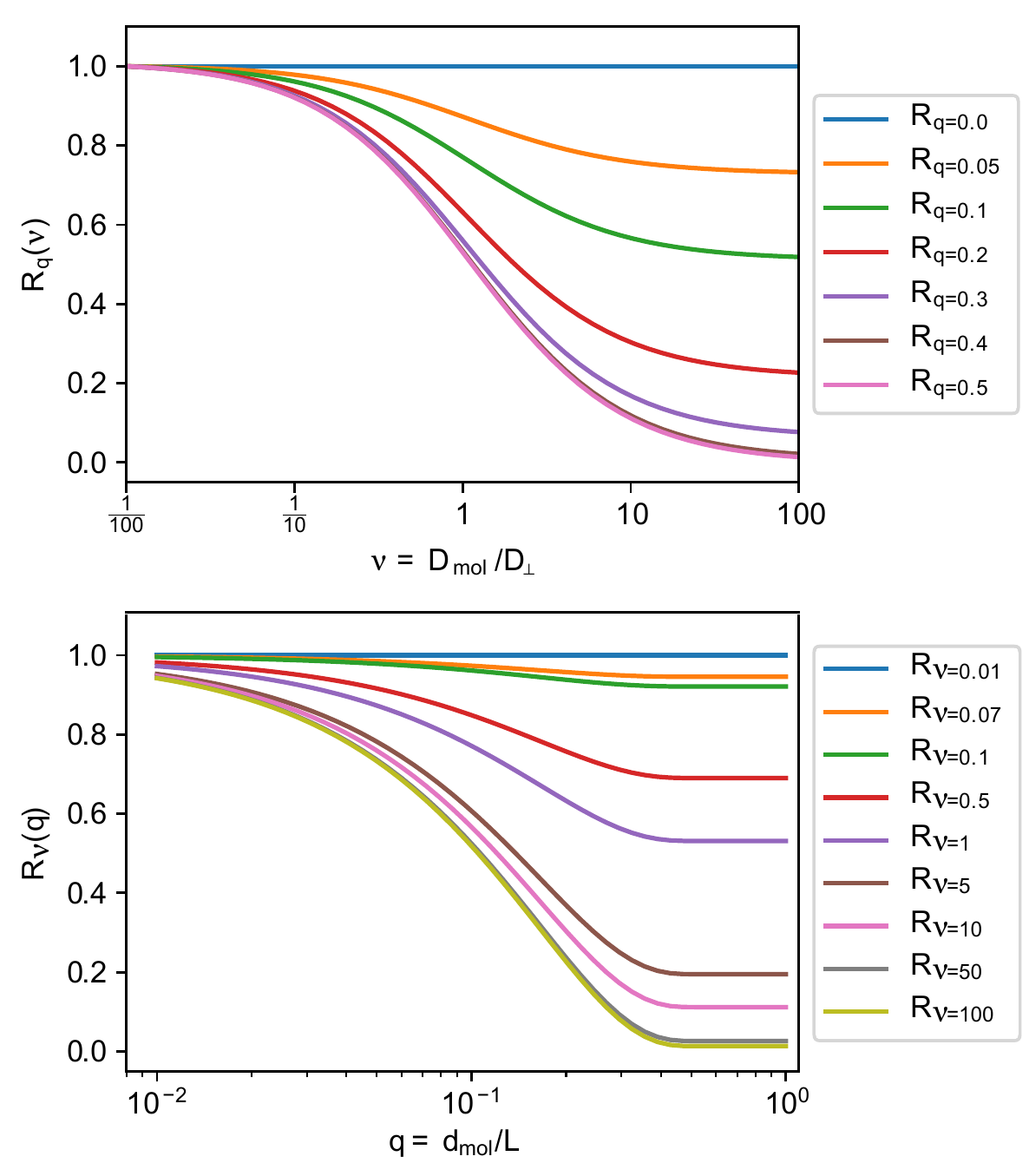}
    \caption{\textbf{Single-parameter plots of the universal correction function $R(\nu,q)$:} Cross sections of the universal correction function $R(\nu,q)$ in the directions of $q$ and $\nu$ in addition to the full surface presented in  \cref{fig:universal_correction_function}}
    \label{fig:app_correction_crossection}
\end{figure}

%
%
%
%
%
%

\section{Fitting procedure for the LWR/EPM method}\label{app:lwr_fit}

As the fit with three fit parameters $d_\mathrm{mol}$, $D_\mathrm{mol}$ and $D_{\perp}$ proved unreliable at approximating the observed data, we opted for a different technique. 

The principal idea is as follows:
\begin{enumerate}
    \item Initially we make a guess for a reasonable value of $D_\mathrm{mol}$. 
    \item For a fixed choice of $D_\mathrm{mol}$ we then perform a fit of $R(\nu, q)$ as a function of $d_\mathrm{mol}$ and $D_{\perp}$ to the values of $$\tilde{\tau}_i=\frac{\tau_\mathrm{EPM,i}}{L_i^2},$$ which provides us with two optimum fit parameters $d_\mathrm{mol}$ and $D_\perp$.
    \item Then we change the value of $D_\mathrm{mol}$ for the fixed values of $d_\mathrm{mol}$ and $D_\perp$ until certain constraints are met and go back to performing the fit for $d_\mathrm{mol}$ and $D_\perp$ until the fit is good enough and the parameters do not change too much.
\end{enumerate}

\paragraph{\textbf{A few more details on the algorithm outlined above:}}

In the case of the ILs presented in this paper, we picked the result of applying the SPM to the where we implicitly assumed  a comparable scale of $D_\mathrm{mol}$ and $D_\perp$.

The switch from $\tau_\mathrm{EPM}$ to $\tilde{\tau}$ (as visualised in \cref{fig:width_reduction_epm_fit}c) removes the scaling with $L$ as predicted by the SPM (as seen in \cref{fig:width_reduction_epm_fit}b) and allows us to focus on the relative deviation of $\tau_\mathrm{EPM}$ from the ideal point-like particle.
It also simplifies visual inspection of the fit results.

The resulting data of $\tilde{\tau}_i$ should -- according to the nature of $R(\nu,q)$ -- exhibit a convergence to a $D_\perp$-dependent value for the limit $L\to\infty$ and a lower convergence limit controlled by $D_\mathrm{mol}$ for $L\to 0$.
This imposes limits on the choice of $D_\perp$ during the fit, because the upper convergence limit needs to be above our maximum measured value for $\tilde{\tau}_i$.
This reduces the options that the fit algorithm has in trying to find the optimum parameters.

In the lower limit $L_i\to 0$, the values of $\tilde{\tau}_i$ should in theory continuously decrease, but the issue of finite time resolution of the simulation steps sets a lower limit for the times $\tau_\mathrm{EPM,i}$ that we can resolve.
This makes the values of $\tilde{\tau}_i$ increase for low values of $L_i$ again after reaching a minimum. 
As this is an artefact of the simulation, we truncate those values of $\tilde{\tau}_i$ for small $L_i$ after attaining their global minimum.

The value of $D_\mathrm{mol}$ for a choice of $d_\mathrm{mol}$ and $D_\perp$ obtained from the fit needs to be chosen such that the lower convergence limit of the fit function lies below the minimum observed value of $\tilde{\tau}_i$. 
We enforce this by iteratively increasing $D_\mathrm{mol}$ if we get a convergence limit that is above the minimum and by decreasing it if the convergence limit is too far below the observed minimum. 
After adapting $D_\mathrm{mol}$, we iterate on the fit of $d_\mathrm{mol}$ and $D_{\perp}$ and the subsequent optimisation of $D_\mathrm{mol}$ until values are converged and the limit constraints are met. 

In our experience, applying this fitting procedure to point-like-particles like described by the SPM (e.g. water) does not yield perfect results, because then there are two excess parameters in the overall fitting process. 
This can easily identified by observing the visual fit results in a representation like \cref{fig:width_reduction_epm_fit}c, where the results for $\tilde{\tau}_i$ will lie on a horizontal line when $d_\mathrm{mol}$ is very small. 
If that is the case, then the SPM should be applied and no attempt at fitting the EPM to those $\tilde{\tau}_i$ results should be made.

%
%
%
%
%
%

\section{Simulation Methods for IL systems} \label{app:simulation_methods}

\begin{figure*}
    \centering
    \includegraphics[width=.85\textwidth]{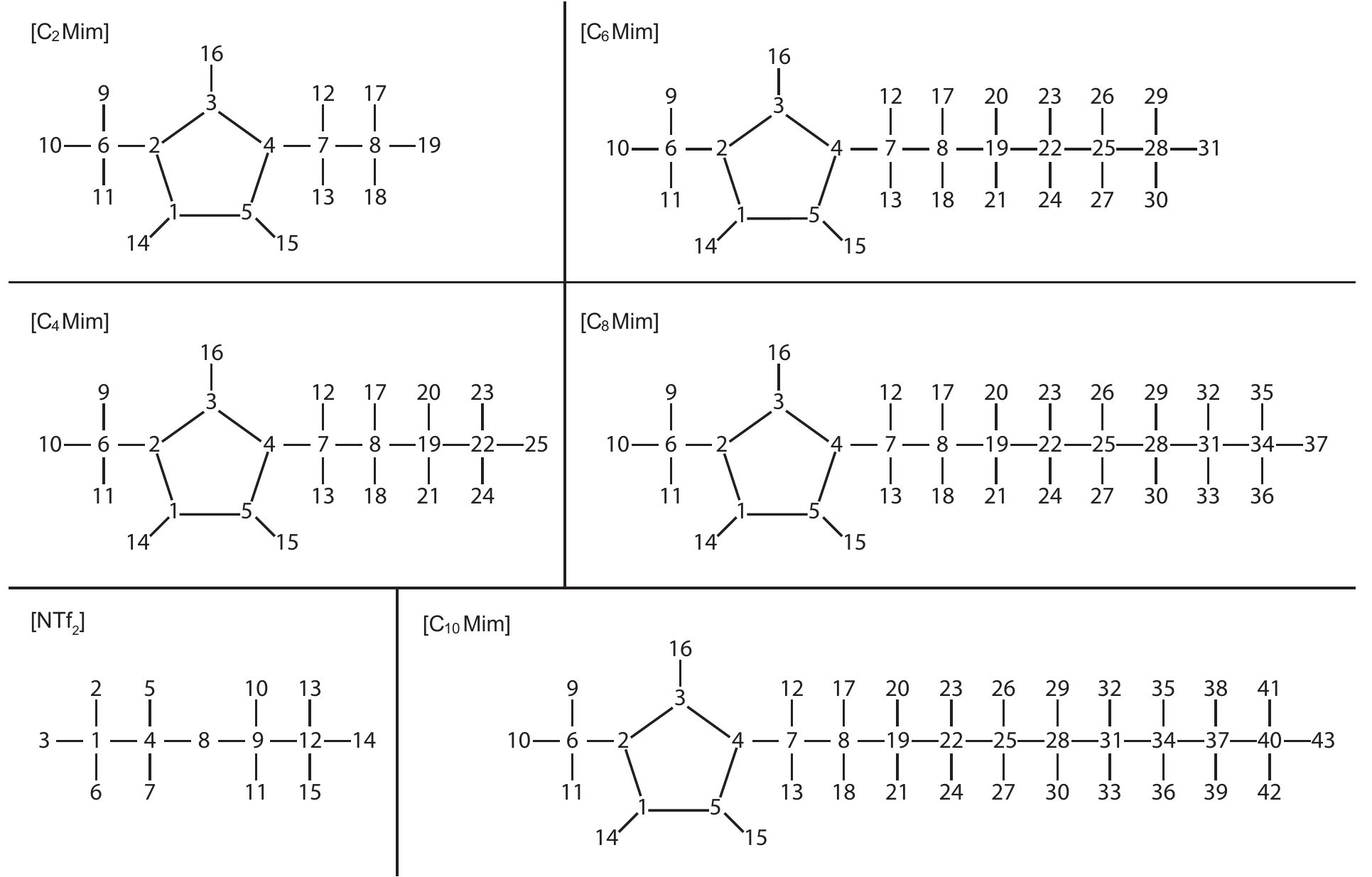}
    \caption{\textbf{Structural information on ILs in MD simulations:} Indices of atoms in the constituting cations [C$_n$Mim]$^+$ ($n = 2,4,6,8,10$) and of the shared anion [NTf$_2$]$^-$. The index corresponds to the row with the associated charge in \cref{tab:force_field_charges}. }
    \label{fig:force_field_indices}
\end{figure*}
\begin{table*}
\centering
    \caption{\textbf{MD simulation information for IL systems:} Table of atomic charges for Imidazolium based cations [C$_n$Mim]$^+$ ($n = 2,4,6,8,10$) and the anion [NTf$_2$]$^-$ obtained by using the Restrained Electrostatic Potential (RESP) method with HF/6–31G(d) level of theory. 
    All presented values are scaled with $0.9$.}
    \label{tab:force_field_charges}
    \begin{tabular}{c|c|c|c|c|c|c}
    \toprule
& [C$_2$Mim]$^+$& [C$_4$Mim]$^+$& [C$_6$Mim]$^+$& [C$_8$Mim]$^+$& [C$_{10}$Mim]$^+$ &[NTf$_2$]$^-$ \\
\midrule
1 & -0.136622 &	-0.14958 &	-0.137267 &	-0.134654 &	-0.132264 &	0.174291\\
2 &	0.0652815 &	0.0900063 &	0.0836334 &	0.0797373 &	0.0780924 &	-0.095428\\
3 &	0.01755 &	 -0.0004824 &	 0.000747 &	0.0011457 &	-0.0014544 &	-0.095428\\
4 &	 0.0226287 &	 0.0170505 &	 0.0097668 & 0.0160227 &	0.0236502 &	1.018366\\
5 &	-0.172285 & -0.15333 & -0.171917 & -0.1769508 & -0.1827252 &	-0.512421 \\
6 &	-0.139399 & -0.148145 & -0.1359441 & -0.1330929 & -0.1333647 &	-0.095428 \\
7 &	 -0.0203112 & -0.0862749 & -0.0532314 & -0.0486414 & -0.048312 &	-0.512421 \\
8 &	 -0.0504603 & -0.055845 & -0.0338382 & -0.0398997 & -0.0366318 &	-0.66306 \\ 
9 &	0.114269 & 0.115376 & 0.112046 & 0.1112931 & 0.1114506 &	1.018366 \\ 
10 &	0.114269 & 0.115376 & 0.112046 & 0.1112931 & 0.1114506 &	-0.512421 \\ 
\midrule
11 &	0.114269 & 0.115376 & 0.112046 & 0.1112931 & 0.1114506 &	-0.512421 \\ 
12 &	  0.0929871 & 0.109707 & 0.0978534 & 0.0943713 & 0.0925002 &	0.174291 \\ 
13 &	  0.0929871 & 0.109707 & 0.0978534 & 0.0943713 & 0.0925002 &	-0.095428 \\ 
14 &	0.215133 & 0.216416 & 0.2133639 & 0.2130768 & 0.212904 &	-0.095428 \\ 
15 &	0.228733 & 0.220349 & 0.2316618 & 0.2330883 & 0.2352042 &	-0.095428 \\ 
16 &	0.200999 & 0.198031 & 0.2031246 & 0.2040003 &	0.2051037	\\ 
17 &  0.0466569 & 0.0566802 & 0.0457668 & 0.0456636 &	0.0437388 \\ 	
18 &  0.0466569 & 0.0566802 & 0.0457668 & 0.0456636 &	0.0437388	\\ 
19 &  0.0466569 & 0.0165366 & -0.0421785 & -0.0497616 &	-0.0468216	 \\ 
20 & &  0.0215739 &  0.0305424 & 0.0356244 &	0.0347736 &	\\
\midrule
21 & &  0.0215739 &  0.0305424 & 0.0356244 &	0.0347736	 \\
22 & &  -0.0600399 & -0.0317277 & -0.0220308 & -0.0230748 &	\\
23 & &  0.0244188 & 0.022545 & 0.0209928 & 0.0204888 &	\\ 
24 & &  0.0244188 & 0.022545 & 0.0209928 & 0.0204888 &	 \\
25 & &  0.0244188 & 0.0091584 & -0.024444 & -0.023118 &	\\
26 & &  &  0.0120222 & 0.0135372 & 0.0130152 &	 \\
27 & &  &  0.0120222 & 0.0135372 & 0.0130152 &	 \\
28 & &  &  -0.0623853 & -0.0123216 & -0.0090108 &	 \\
29 & &  &  0.0211455 & 0.0061128 & 0.0067356 &	 \\
30 & &  &  0.0211455 & 0.0061128 & 0.0067356 &	 \\
\midrule
31 & &  &  0.0211455 & 0.0376824 & 0.0020568 &	 \\
32 & &  &  & -0.0022692 & 0.0013164 &	 \\
33 & &  &  & -0.0022692 & 0.0013164 &	 \\
34 & &  &  & -0.048024 & -0.0036876 &	\\
35 & &  &  & 0.0143748 & 0.0058968 &	\\
36 & &  &  & 0.0143748 & 0.0058968 &	 \\
37 & &  &  & 0.0143748 & 0.0187032 &	\\
38 & &  &  &  & 0.0035028 &	\\
39 & &  &  &  & 0.0035028 &	\\
40 & &  &  &  & -0.060936 &	 \\
\midrule
41 & &  &  &  & 0.0157992 &	\\
42 & &  &  &  & 0.0157992 &	\\
43 & &  &  &  & 0.0157992 &	\\
\bottomrule
    \end{tabular}
\end{table*}

The analysis of complex IL systems is based on MD simulations of [C$_n$Mim][NTf$_2$] ($n = 2,4,6,8,10$) which were performed in GROMACS 5.1.2 \cite{van2005gromacs}  with a $\SI{2}{\femto\second}$ time step. 
For benchmarking we used simulations of bulk ILs ($1000$ ion pairs, $\SI{100}{\nano\second}$ production run), for the studies of supported IL films, we use the same ILs ($1800$ ion pairs, $\SI{200}{\nano\second}$ production runs), deposited on a neutral hydroxylated alumina surface. 
All simulations were performed as discussed in detail in our previous work \cite{vucemilovic2019insights}.

In short, the Van der Waals parameters for [C$_n$Mim]$^+$ ($n = 2,4,6,8,10$) cations and [NTf$_2$]$^-$ were adapted from Maginn \cite{kelkar2007effect} and Canongia Lopez and Padua \cite{canongia2004modeling}, respectively. 
In all systems, the partial atomic charges 
were derived from quantum mechanical calculations (HF/6–31G level of theory) and were rescaled to $\SI{90}{\percent}$ of their original values \cite{cornell2002application,vucemilovic2019insights}.
 
See \cref{fig:force_field_indices} for atom indices and \cref{tab:force_field_charges} for resulting charges of the respective atoms.

Van-der-Waals and Coulomb interactions were cut-off at $\SI{1.2}{\nm}$.
In all production IL simulations, the temperature is controlled with the Nosé-Hoover thermostat \cite{nose1984unified,hoover1985canonical} with a coupling time of \SI{0.4}{\ps}, while for the water system the BDP velocity rescaling thermostat \cite{Bussi2007} is used with a coupling time of \SI{0.5}{\ps}.
The IL systems are created initially with random configurations of the solvent placed either in a cubic box (pure liquid system, \cref{fig:simulation_samples}) or a triclinic box (system with a solid \cref{fig:scheme_system}).
In the system with a solid, this results in a slit pore due to the periodic boundary conditions that are applied in all three systems.

\subsection{Pure IL system}
After minimising the energy of the system to create a numerically stable starting configuration, the system was relaxed in an NVT ensemble for \SI{5}{\ns}.
Afterwards a simulated annealing step was conducted in the NPT ensemble for $\SI{12}{\nano\second}$ ($P =\SI{1}{\atmospheric}$, $\beta = \SI{4.8e-5}{\per\mega\pascal}$) employing the Berendsen barostat \cite{berendsen1984molecular}.
The system was heated from \SI{300}{\K} to \SI{700}{\K} over \SI{3}{\ns} and kept there for another \SI{3}{\ns}.
It was then cooled down back from \SI{700}{\K} to \SI{300}{\K} over \SI{4}{\ns}, where it was kept for another \SI{2}{\ns} with the Nosé-Hoover thermostat to allow for final adjustments of the density to atmospheric pressure.
The box scaling was isotropic.
The obtained model system was simulated via the NVT ensemble (cubic box) at $T=\SI{300}{\K}$ for \SI{100}{\ns} as the production run, with the temperature again being controlled via the Nosé-Hoover thermostat.

\subsection{S-L-V IL system}
The simulated S-L-V system contained of a slab of sapphire ($\SI{7.57}{\nano\meter}\times \SI{6.29}{\nano\meter} \times  \SI{2.12}{\nano\meter}$) optimised in GULP \cite{gale2003general} with a fully hydroxylated ($0001$) $x$-$y$ surface, defined by the CLAYFF \cite{cygan2004molecular} force field.
We positioned $1800$ ion pairs of IL above the surface of the sapphire (\SI{2.1}{\nano\meter}) in a monoclinic simulation box.
Coupling between the IL and the solid surface was performed using the Lorentz-Berthelot mixing rules \cite{brkljaca2015complementary}.
The systems were first minimised and then semi-isotropic NPT annealing simulations, using 
$P =\SI{1}{\atmospheric}$, $\beta = \SI{4.8e-5}{\per\mega\pascal}$, were performed for $\SI{12}{\nano\second}$, the subsequent annealing procedure being the same as for the pure IL system.
The box scaling was only allowed to vary in the $z$-direction orthogonal to the solid-liquid interface.
Afterwards an $\SI{80}{\nano\meter}$ large vacuum slab is placed on top of the solid-liquid film with the purpose of decreasing the contribution of the z-replicas to the electrostatic interactions in the central simulation cell, thus creating a solid-liquid-vacuum system (see \cref{fig:app_slv_system_sim}). 
The model system then spans a height of about \SI{100}{\nano\meter} in the $z$-direction.
The obtained model system was simulated via the NVT ensemble at $T=\SI{300}{\K}$ for \SI{200}{\ns} as the production run, with the temperature again being controlled via the Nosé-Hoover thermostat using the Langevin algorithm \cite{nose1984unified,hoover1985canonical}.

\begin{figure}
    \centering
    \includegraphics[width=.48\textwidth]{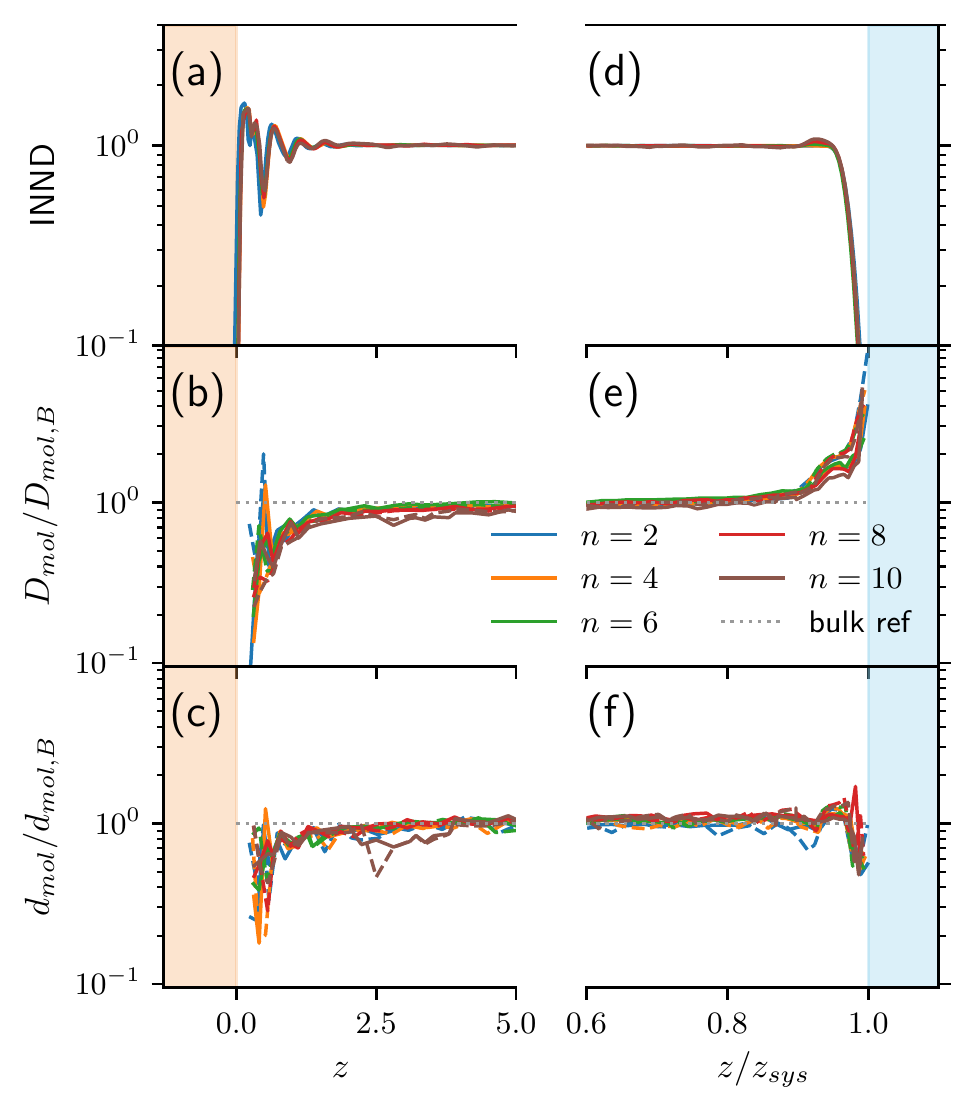}
    \caption{\textbf{Deformation parameters extracted from IL SLV systems:} Total added up INND of ILs at the (a) solid and (d) vacuum interfaces relative to INND in bulk.
    $D_\mathrm{mol}$ profiles relative to the bulk system value at the (b) SL and the (e) LV interfaces.
    $d_\mathrm{mol}$ profiles relative to the bulk system value at the (c) SL and the (f) LV interfaces.
    In $D_\mathrm{mol}$ at the LV (e), we see profiles similar to $D_\perp$ in \cref{fig:IL_slv_innd_diffusion_profiles}h, whereas the behaviour at the SL interface (c) is much closer to that of $D_\parallel$ in \cref{fig:IL_slv_innd_diffusion_profiles}d.
    Overall, $D_\mathrm{mol}$ drops to about a quarter of its bulk value at the SL (b) and climbs to about $4$ times its bulk value at the LV interface (e).
    Also $d_\mathrm{mol}$ is only affected at the very contact layers close to the interfaces, where deformation in constrained by the interface being present. 
    More specifically, $d_\mathrm{mol}$ exhibits a trend of dropping to about half the bulk value at the SL interface (c) as well as at the LV interface (f) with the variance being higher at the SL.}
    \label{fig:app_deformation_IL_SLV}
\end{figure}

%
%
%
%
%
%
\section{Error estimates}\label{app:error_estimates}
In our paper, we use several differently distributed random variable statistics. 
Most notable among those are the Mean Square Displacement (MSD) of the Einstein approach.
We generally assume the MSD to be the estimated variance of a normal distributed random variable. 
Hence, the MSD is a prime example of a $\chi^2$-distributed variable for which we can provide an error estimator via the usual estimator for the variance. 

Let 
\begin{align}
    S= \frac{1}{n-1}\sum_n (X_n-\overline{X})^2
\end{align}
denote the standard unbiased estimator for the variance of a sample set of size $n$. 
Then 
$$Q=\frac{(n-1)S}{\sigma^2}$$
is expected to be $\chi^2_{n-1}$-distributed ($\chi^2$ with $n-1$ degrees of freedom) and we can use this to derive a $(1-\alpha)100\si{\percent}$-confidence interval for $\sigma^2$:

\begin{align}
    P\left(\sigma^2 \in \left[\frac{(n-1)S^2}{\chi^2_{\frac{\alpha}{2},n-1}},\frac{(n-1)S^2}{\chi^2_{1-\frac{\alpha}{2},n-1}}\right]\right) = 1-\alpha
\end{align}

where $\chi^2_{p,k}$ is defined by:
\begin{align}
    P(X > \chi^2_{p,k}) = 1-p
\end{align}
with $X$ being a $\chi^2_{k}$-distributed random variable.

Wherever an error or confidence interval for the MSD is denoted in our graphs or calculations (e.g. in linear fits for the derivation of $D$), we use this estimate of the confidence interval with $\alpha=0.05$.

For the lifetime distributions, our calculations show, that the lifetimes are distributed like an overlay of multiple exponential functions. 
To obtain a confidence interval for the mean lifetime, we thus use the mid- to long-term approximation of the lifetime being approximately exponentially distributed to derive a confidence interval.

Let $\overline{X}$ denote the mean lifetime of a sample set obtained from $n$ data points, then the $(1-\alpha)100\si{\percent}$-confidence interval for the mean lifetime $\tau$ is given by:
\begin{align}
    P\left(\tau \in \left[\frac{2n\overline{X}}{\chi^2_{\frac{\alpha}{2},2n}},\frac{2n\overline{X}}{\chi^2_{1-\frac{\alpha}{2},2n}}\right]\right) = 1-\alpha
\end{align}
Hence, this sets a confidence interval to estimate the true range of $\tau$ with $\alpha=0.05$ as well as the estimate of the resulting error of the mean diffusion coefficient $D_\perp$.

\begin{figure*}
    \centering
    \includegraphics[width=.8\textwidth]{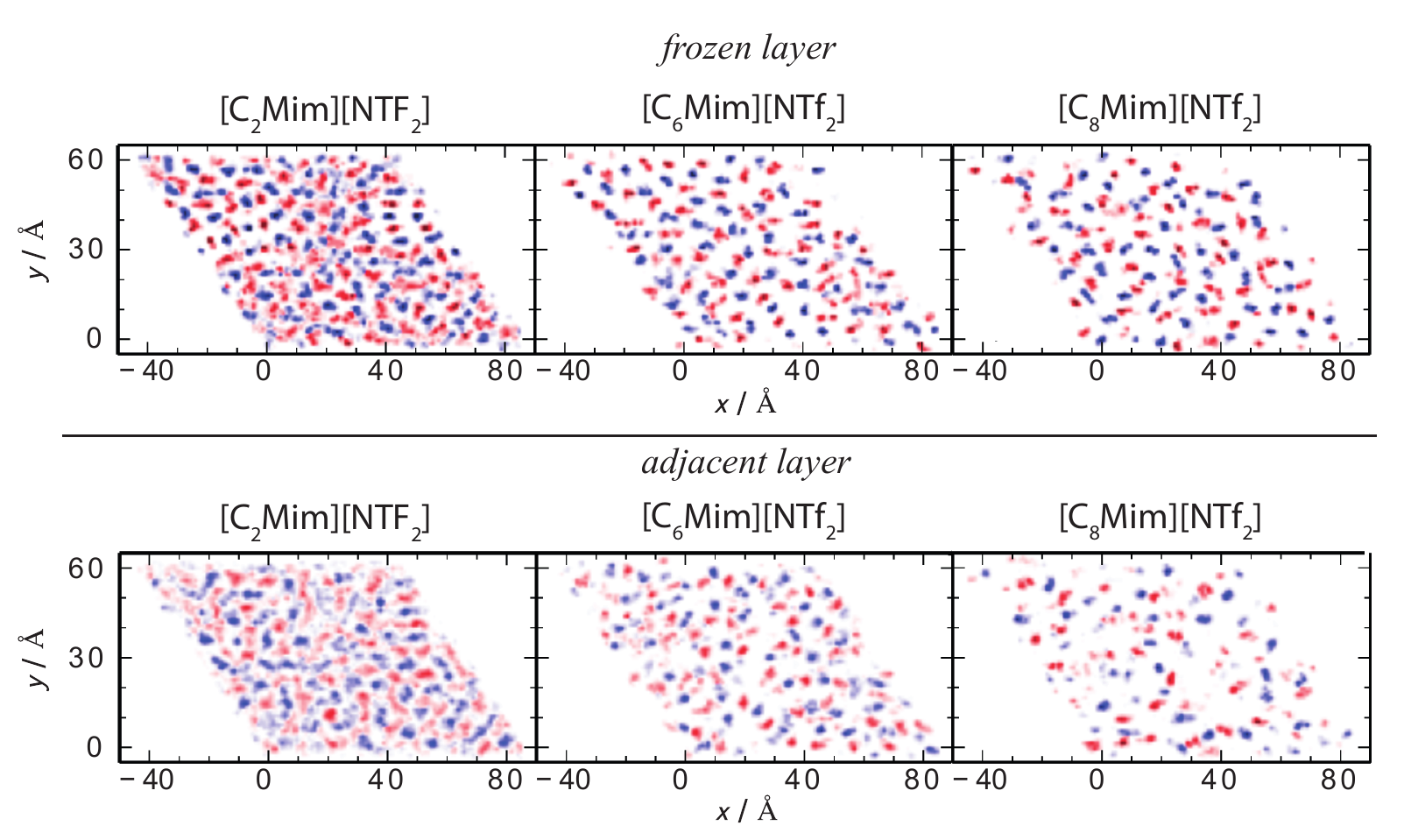}
    \caption{\textbf{Visualisation of the frozen layer on top of the solid in IL SLV systems:} Residence probability distribution of anions (blue) and cations (red) in the frozen layer (first \SI{1}{\angstrom} from the SL interface) and the adjacent layer (next \SI{5}{\angstrom}) for $n=2,4,8$ in addition to the systems presented in  \cref{fig:frozen_layer_c4_c10}. \textit{(Figure adapted from \cite{vuvcemilovic2021computational})}}
    \label{fig:frozen_layer_c2_c6_c8}
\end{figure*}

%
%
%
%
%
%

\section{Experimental reference values}\label{app:rtils_experimental}
We used experimental results from Tokuda et al. \cite{tokuda2005physicochemical} to provide an experimental reference value for our simulation-based diffusion results for $n=2,4,6,8$ (no experimental results for $n=10$ were available to us). 
In the referenced work, Tokuda et al. provide per-particle-family (i.e. cation and anion) and temperature-dependent results for the self-diffusion coefficient $D$ in the form 
\begin{align}
    D(T)= D_0 \exp\left(-\frac{B}{T-T_0}\right),
\end{align}
where they provide the values of $D_0$, $B$ and $T_0$ including respective error estimates ($\Delta D_0$, $\Delta B$ and $\Delta T_0$) in a tabular fashion. 
As our simulations are run at $T=\SI{300}{\kelvin}$, we used their values for the parameters and their error estimates to calculate the values of $D(T)$ and a respective error estimate $\Delta D(T)$ via the formula:
\begin{align}
    \Delta D(T) =& \left(\Delta D_0 + D_0\frac{\Delta B \times (T-T_0) + B \times \Delta T_0}{(T-T_0)^2} \right)\notag\\
    &\times\exp\left(-\frac{B}{T-T_0}\right).
\end{align}
In \cref{tab:correction_model_benchmark_spm}, we denote the obtained values and error estimates as $D(T) \pm \Delta D(T)$.

%
%
%
%
%
%

\section{Considerations for the choice of $L$}\label{app:choice_of_L}

The results obtained from a trajectory analysis using the SPM and EPM/LWR methods depend on the time-resolution of the trajectory.
Since the time difference between subsequent frames limits the statistical resolution of short lifetimes as a consequence of discretisation, the choice of simulation settings -- most notably the total simulated duration and the time difference between subsequent frames -- is essential for lower and upper bounds on viable slice thicknesses $L$.
Additionally, small-size effects in very small bulk systems can cause anisotropy to arise, which may also deviate from assumptions of our model, thus reducing its accuracy \cite{yeh2004system}. 
At small values of $L$, the very short expected and observed lifetimes can also be too short to actually resolve the diffusive regime setting an absolute lower limit to possible slice thicknesses that even shorter simulation time steps cannot reduce further. 

Overall, we wish to point out, that the choice of $L$ and accompanying simulation parameters must be picked carefully or one may end up with parameters for the wrong particle dynamics or simply discretisation artefacts.

%
%
%
%
%
%

\section{Additional software resources}\label{app:provided_software}
We supply further material on github and Zenodo. Most notably, the accompanying github project page \url{https://github.com/puls-group/diffusion_in_slit_pores} contains the pre-calculated tabular data of the universal correction function for bulk-like slabs $R_B(\nu,q)$ as well as the script to calculate $R_B(\nu,q)$ in general via a numerical solution to the underlying two-dimensional Smoluchowski equation. 
In addition to the github project, where active development may be going on and where we also invite feedback and bug reports, we also offer an archived version of the code on Zenodo \cite{hollring_kevin_2022_7446071}, where the latest version of the project has been archived to ensure long-time availability and a clearly documented state at time of publication.

\FloatBarrier

\bibliographystyle{model1a-num-names}


\bibliography{diffusion_jcis_epm}



\end{document}